\documentclass[twocolumn,showpacs]{revtex4}
\usepackage{times,xspace}
\usepackage{amsbsy,amssymb,amsmath,bm}
\usepackage{graphicx,color,epsfig,rotate}
\usepackage{fancyheadings}

\def\one{{\mathchoice {\rm 1\mskip-4mu l} {\rm 1\mskip-4mu l} {\rm
          1\mskip-4.5mu l} {\rm 1\mskip-5mu l}}}
\def\bbbc{{\mathchoice {\setbox0=\hbox{$\displaystyle\rm C$}\hbox{\hbox
            to0pt{\kern0.4\wd0\vrule height0.9\ht0\hss}\box0}}
            {\setbox0=\hbox{$\textstyle\rm C$}\hbox{\hbox
            to0pt{\kern0.4\wd0\vrule height0.9\ht0\hss}\box0}}
            {\setbox0=\hbox{$\scriptstyle\rm C$}\hbox{\hbox
            to0pt{\kern0.4\wd0\vrule height0.9\ht0\hss}\box0}}
            {\setbox0=\hbox{$\scriptscriptstyle\rm C$}\hbox{\hbox
            to0pt{\kern0.4\wd0\vrule height0.9\ht0\hss}\box0}}}}

\newcommand{\greenf}[1]{\langle\!\langle{#1}\rangle\!\rangle}

\newcommand{\ignore}[1]{}
\newcommand{\lComment}[1]{}
\newcommand{\mComment}[1]{}
\newcommand{\gComment}[1]{}
\renewcommand{\lComment}[1]{\textcolor{blue}{Lev: #1}}
\renewcommand{\mComment}[1]{\textcolor{green}{Marina: #1}}
\renewcommand{\gComment}[1]{\textcolor{red}{Gerardo: #1}}
\begin{document}
\title{Tunneling measurement of quantum spin oscillations}
\author{L.N. Bulaevskii$^1$, M. Hru\v{s}ka  $^{1,2}$, and G. Ortiz$^1$}

\affiliation{$^1$ Theoretical Division,
            Los Alamos National Laboratory, Los Alamos, NM 87545}
\affiliation{$^2$ Department of Physics, University of Washington,
            Seattle, WA 98195}
\date{Received \today }

\begin{abstract}
	    We consider the problem of tunneling between two leads via
a localized spin 1/2 or any other microscopic system (e.g., a quantum
dot) which can be modeled by a two-level Hamiltonian.  We assume that a
constant magnetic field ${\bf B}_0$ acts on the spin, that electrons in
the leads are  in a voltage driven thermal equilibrium and that the
tunneling electrons are coupled to the spin through exchange and
spin-orbit interactions. Using the non-equilibrium Keldysh formalism we
find the dependence of the spin-spin and current-current  correlation
functions on the applied voltage between leads $V$, temperature $T$,
${\bf B}_0$, and on the degree  and orientation ${\bf m}_{\alpha}$ of
spin polarization of the electrons in the right  ($\alpha=$R)  and left
($\alpha=$L) leads. We show that  a) The spin-spin correlation
function  exhibits a peak  at the Larmor frequency, $\omega_L$,
corresponding to the effective magnetic field ${\bf B}$  acting upon
the spin  as determined by ${\bf B}_0$ and the exchange field induced
by tunneling of spin polarized electrons, b)  If the ${\bf
m}_{\alpha}$'s are not parallel to ${\bf B}$ the second order
derivative of the average tunneling  current $I(V)$  with respect to
$V$ is proportional to the spectral density of the spin-spin
correlation function,  i.e., exhibits a  peak at the voltage
$V=\hbar\omega_L/e$, c) In the same situation when $V>B$  the
current-current correlation function exhibits a peak at the same
frequency, d) The signal-to-noise (shot noise) ratio $R$ for this peak 
reaches a maximum value of order unity, $R\le 4$, at large $V$  when
the spin is decoupled from the  environment and the electrons in both 
leads are fully polarized in the direction perpendicular to ${\bf B}$, 
and e) $R\ll 1$ if the electrons are weakly polarized, or if they are
polarized in a direction close  to ${\bf B}_0$, or if the spin
interacts with the environment stronger than with the tunneling
electrons.  Our results of a full quantum-mechanical treatment of the
tunneling-via-spin model when $V \gg B$ are in  agreement with those
previously obtained in the quasi-classical approach. We discuss also
the experimental results observed using STM dynamic probes of the
localized spin.

\end{abstract}

\pacs{03.65.Xp, 03.65.Ta, 73.40.Gk, 73.63.Kv}

\maketitle

\section{Introduction}
\label{sec1}

            Interest in quantum information processing has brought
significant attention to the problem of measurement of tunneling
currents via a microscopic system that can be modeled by a two-level
Hamiltonian (such as a quantum dot, or a molecule or an atom with a
localized spin
\cite{BulaevskiiOrtiz,ParcolletHooley,KorotkovAverin,RuskovKorotkov}).
In the case of a single spin, measurements of the tunneling current in
such a system provide information on spin orientation and its dynamics,
and they constitute an example of indirect-continuous quantum
measurement \cite{BulaevskiiOrtiz}. A fundamental question that arises
is what signatures of the spin dynamics are encoded in the tunneling
current and how this current affects the spin dynamics.

Scanning tunneling microscopy (STM) experiments \cite{Manassen,Durkan} 
on a single molecule  with a spin, in the presence of a magnetic field
${\bf B}_0$, have reported a peak in the current noise power
spectrum  (i.e., current-current correlation function) ${\cal
P}(\omega)$  at the Larmor frequency $\omega_L=\gamma B_0$, where
$\gamma$ is the  gyromagnetic ratio. Experiments were done at room
temperature and the authors found that the signal-to-noise ratio $R$
(ratio of the power at the peak frequency to the shot noise power)
exceeded  unity and was almost independent of the orientation of the
applied magnetic field  ${\bf B}_0$. In the non-relativistic approach
the tunneling electrons  couple to the spin by exchange interaction. In
this case, electrons with a spin polarization along ${\bf B}_0$ do not
couple with the oscillatory components of the spin (which  are
perpendicular to ${\bf B}_0$). The experimental results
\cite{Manassen,Durkan}  are difficult to explain in the framework of a
single-spin non-relativistic model  since electrons in the leads were
polarized by the same magnetic field which acted on the spin.  Possible
relevance of spin-orbit interaction to explain these data was discussed
by  Shachal and Manassen and later on by Balatsky and Martin
\cite{Balatsky}. Recently, Levitov and Rashba \cite{LevitovRashba}
noticed that  in systems with low space symmetry (as in dots or a
molecule near a surface) the nonvanishing orbital moment of electrons
near the spin (in the dot or molecule)  provides a  strong coupling of
the tunneling electrons to the spin, and they speculated that  this
mechanism  may lead to a significant effect of the spin oscillatory
component on the  tunneling current.  In fact, in order to fit the
experimental data \cite{Manassen,Durkan} one needs  a model of electron
tunneling via a single spin which is able to explain  not only the
existence of a peak in the current power spectrum,  but also a
significant  signal-to-noise ratio $R>1$ almost independent of the
orientation of ${\bf B}_0$.

In the following we consider a  model with exchange (non-relativistic)
coupling of  a single spin $1/2$  and  the tunneling electrons,
including spin-orbit coupling. In the  framework of  this model we
analyze the dependence of $R$ and linewidth $\Gamma$ on the applied
voltage $V$  between leads, the applied magnetic field ${\bf B}_0$, the
temperature $T$, and on  the degree and orientation ${\bf m}_{\alpha
0}$ of electron polarization in the right ($\alpha=$R)  and left
($\alpha=$L) leads in the steady state. This state establishes in a
transient time after the voltage or tunneling matrix elements are
switched on.

The problem of tunneling via a single localized spin $1/2$ is similar
to the problem  of tunneling via a double quantum dot (two-level
system) extensively studied by  Korotkov and Averin
\cite{KorotkovAverin} and by Ruskov and Korotkov
\cite{RuskovKorotkov}.  Indeed, they were interested in the question of
how quantum oscillations in the two-level  system, representing a
qubit, may be detected by a tunneling current.  Such a system can be
described by a model of electron tunneling via a spin when leads
electrons are fully polarized.  Using the Bloch equations describing
the ensemble-averaged evolution of the density matrix for the coupled
qubit-detector system, they obtained oscillations at the Larmor
frequency in the  noise power spectrum ${\cal P}(\omega)$, with $R\leq
4$ in the case of weak interaction between the qubit and the detector.
A similar quasi-classical approach for tunneling of electrons via a
single spin was used in Ref.~\cite{BulaevskiiOrtiz}. In such an
approximation, the action of the electrons on the spin is replaced by
the action of an effective {\it classical} magnetic field with a shot
noise spectrum depending on the tunneling current between the leads.
This approach has sense at high currents (voltages), while at low
currents it does not account for the quantum nature of the tunneling
electrons and hence a  more elaborate solution of the quantum transport
equation for both spin and electrons is desired. Here the situation is
formally similar to the case of the photo-electric effect where in some
circumstances a full quantum mechanical treatment of the problem
provides the same information as a semi-classical approach, while in
other cases, such as the measurement of the statistics of the
photo-current fluctuations, the predictions of the two approaches
differ qualitatively \cite{FunkBeck}.

	    An important step toward a full quantum mechanical
treatment of electron tunneling via a spin was recently given by
Parcollet and Hooley \cite{ParcolletHooley}. Assuming that  the spin
interacts only with the tunneling electrons, and electrons in the leads
are  in a voltage driven thermal equilibrium, they computed the spin
magnetization of a quantum dot in the two-leads Kondo model as a
function of the temperature of the electrons, magnetic field, and
voltage between the leads (which exceeds the Kondo temperature $T_K$)
in the steady state, i.e., in the long time limit.  They considered the
situation where electrons in the leads are polarized in the same
direction  as the magnetic field acting on the spin, ${\bf m}_{{\rm
R}0}={\bf m}_{{\rm L}0}={\bf m}_0$, ${\bf m}_0\parallel {\bf B}_0$.
Their calculations confirmed previously  obtained quasi-classical
result \cite{Glazman} that, in the steady state even at zero order in
the spin-leads coupling, the magnetization is not given by the thermal
equilibrium expression $M_{\sf eq}=(1/2)\tanh(B_0/2T)$ when the spin is
more strongly coupled to the leads than to any other thermal bath. At
long times, the state of the spin is completely determined by the
characteristics of the tunneling electrons, and since the system is
out  of equilibrium, the steady state thus achieved is described by a
steady state distribution function which differs from the Gibbs
distribution.  It turns out that this distribution function depends
upon the voltage between the leads. They also showed that the correct
way to calculate the perturbative corrections to the spin distribution
function and the spin decoherence rate  $\Gamma$ in the steady state is
by solving a quantum transport equation self-consistently.  We will
extend the treatment of Parcollet and Hooley to arbitrary orientations
of ${\bf m}_{{\rm R}0}$,  and ${\bf m}_{{\rm L}0}$ with respect to
${\bf B}_0$, and  account for a direct tunneling of electrons between
leads.

            The main goals in this article are (i) to examine under
what tunneling conditions  the  current-current correlation function
${\cal P}(\omega)$ exhibits a peak at the Larmor frequency in the
steady state, and  (ii) to calculate $\Gamma(V,T,{\bf B}_0,{\bf
m}_{{\alpha} 0})$ and $R(V,T,{\bf B}_0,{\bf m}_{{\alpha} 0})$ in the
regime where $eV \gg T_K$,  with $e$ denoting the charge of the
electron.  As in Ref.~\cite{ParcolletHooley} we use the Keldysh
formalism \cite{Schwinger,Keldysh} and the Majorana-fermion
representation \cite{Tsvelik} for the spin to find the spin
distribution function and the current-current correlation function.

	    We note, from a more general perspective, that the problem
of electron tunneling via a spin represents an example of indirect
quantum measurement. The spin is probed by the tunneling electrons
whose correlation function is measured in a continuous fashion with a
classical apparatus (which is not included). The goal of quantum
measurement is to determine the frequency of precession of the isolated
spin (energy separation in the isolated two-level system) and to obtain
information on the initial state of the system. However, when the spin
is decoupled from the environment, tunneling via the spin changes its
state as the measurement goes on. After switching on the voltage, or
the tunneling matrix elements, there is an initial period of time where
the tunneling current depends upon the initial state of the spin. After
some transient time the steady state, which is in general independent
of the initial conditions, establishes. In the steady state the
frequency of precession is renormalized by the tunneling electrons, and
the width of the peak (at the precession frequency) and corresponding
signal-to-noise ratio depend upon the precession frequency and
characteristics of the tunneling electrons. To extract information
about the precession of the isolated spin one needs to have complete
information on the tunneling electrons. We show how to relate the
precession frequency of the isolated spin to results of current
measurements in the steady state (peak frequency, width of the peak and
signal-to-noise ratio). To obtain information on the initial  state of
the spin, measurements of the current during the initial transient
period are needed.  Here we do not consider the spin dynamics in this
short time interval.

The plan of the article is the following. In the next section we
present the model for tunneling via a single spin 1/2, with and without
spin-orbit coupling, and a model describing tunneling via a two-level
quantum dot. Then we introduce Majorana fermions and the Keldysh
technique. In the subsequent section we will compute the spin
distribution function, broadening of the spin precession due to the
tunneling of electrons, and the spin-spin correlation function. Then we
will calculate the dependence of the tunneling current on the spin and
determine the current-current correlation function ${\cal P}(\omega)$.
Finally, we discuss the STM experimental results and compare our
results to those obtained in the quasiclassical approach
\cite{BulaevskiiOrtiz,RuskovKorotkov,KorotkovAverin} and other
theoretical works \cite{Balatsky,LevitovRashba,BalatskyManassen}.

\section{The model}
\label{sec2}

\begin{figure}
\centerline{\epsfxsize=6.0cm \hspace*{-0.7cm} \epsfbox{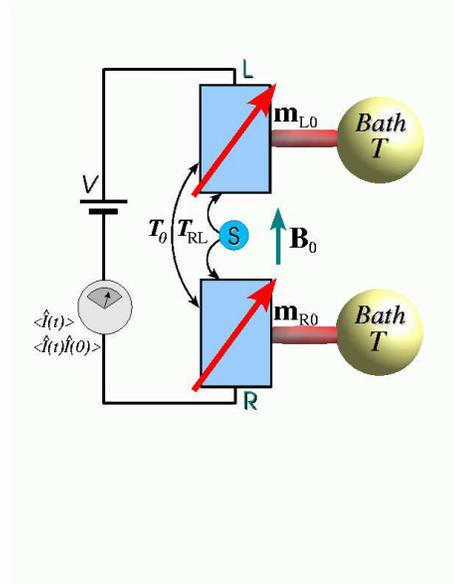}}
\caption{Schematics of the physical systems represented by the
Hamiltonian $\cal H$, and measurement process involved in determining
the current $\langle \hat{I}(t)\rangle$ and current-current correlation
function  $\langle \hat{I}(t)\hat{I}(0)\rangle$ which contain encoded
information about the spin $\bf S$. The electronic tunneling current is
established by a dc  voltage $V$.
}
\label{fig1}
\end{figure}

For the system consisting of a spin coupled to the leads by
an exchange mechanism (in the non-relativistic approximation, i.e.,
neglecting the spin-orbit interaction), we  use the Hamiltonian of the
two-leads Kondo model \cite{BulaevskiiOrtiz,ParcolletHooley} with a
direct tunneling term included (which we call tunneling-via-spin (TvS)
model)
\begin{eqnarray}
{\cal H}&=&{\cal H}_e+{\cal H}_s+{\cal H}_T, \ \ \
{\cal H}_T={\cal H}_{\sf ref} + {\cal H}_{\sf tr}, \label{Ham} \\
{\cal H}_e&=&\!\!\!\sum_{\alpha, n, \sigma,\sigma'}\!
[\epsilon^{\;}_{n \alpha}
\delta_{\sigma\sigma'}-\frac{1}{2}{\bf B}_{\alpha} \cdot
\vec{\bm{\sigma}}_{\sigma\sigma'}] c^{\dagger}_{\alpha n
\sigma}c^{\;}_{\alpha n \sigma'}, \nonumber \\
{\cal H}_s&=&-g\mu_B {\bf B}_0\cdot{\bf S}, \nonumber \\
{\cal H}_{\sf ref}&=&\!\!\!\!
\sum_{\alpha,n,n' \sigma, \sigma'}\!\!\!\!\!
c^{\dagger}_{\alpha n \sigma}(\hat{T}_{\alpha\alpha})_{\sigma\sigma'}
c^{\;}_{\alpha n' \sigma' } , \ \ \
\hat{T}_{\alpha\alpha}=T_{\alpha\alpha}^{({\sf ex})}{\bf
S}\cdot\vec{\bm{\sigma}}_{\sigma\sigma'} ,
\nonumber \\
{\cal H}_{{\sf tr}}&=&\!\!\!\!\sum_{n,n',
\sigma,\sigma'} \!\!\!\! c^{\dagger}_{{\rm R} n \sigma}
(\hat{T}_{{\rm RL}})_{\sigma\sigma'}
c^{\;}_{{\rm L} n' \sigma' }\! + {\rm H. c.} ,
\nonumber \\
&&(\hat{T}_{{\rm RL}})_{\sigma\sigma'}= T_0\delta _{\sigma\sigma'}+ T_{{\rm RL}}^{({\sf ex})}{\bf S} \cdot
\vec{\bm{\sigma}}_{\sigma\sigma'} ,
\nonumber
\end{eqnarray}
where $c^{\dagger}_{\alpha n \sigma}$ ($c^{\;}_{\alpha n
\sigma}$) creates (annihilates) an electron in the left or right lead
(depending on $\alpha \in \{{\rm L,R}\}$) in the eigenstate $n$, and
with spin $\sigma$. Further, $\epsilon^{\;}_{n \alpha}=\epsilon^{\;}_n
-\mu_{\alpha}$,  where $\epsilon_n$ is the energy in the state $n$ and
$\mu_{\alpha}$ is the  chemical potential in the lead $\alpha$, while
$\vec{\bm{\sigma}}$ represents the three Pauli matrices. $T_{{\rm
LL}}^{({\sf ex})}$, $T_{{\rm RR}}^{({\sf ex})}$ and  $T_{{\rm
LR}}^{({\sf ex})}$ are tunneling matrix elements due to the exchange
interaction  for the electron tunneling from the leads to the molecule
with the spin 1/2, while  $T_0$ is the direct tunneling matrix element.
We take them as real numbers.  The spin localized in the molecule is
described by the operator ${\bf S}=(S_x,S_y,S_z)$. Figure \ref{fig1}
sketches the physical setup we want to study and which basically
represents the model Hamiltonian $\cal H$.

In the following we use energy units for the bias voltage $V$, i.e., we
denote $eV$ by $V$. Also, by $B$ we mean $g\mu_B B$, write $T$ instead
of $k_B T$, and  $\omega$ represents $\hbar\omega$.

We describe the leads by a free electron gas with a density of states
$\rho(\epsilon)$ of bandwidth $D$, so that $\epsilon_n$ is the bare
energy of the electron in the eigenstate $n$, the same in both leads.
We assume weak tunneling, $T_0^2\rho_0^2,|T_{{\rm RL}}^{({\sf
ex})}|^2\rho_0^2\ll 1$, where $\rho_0$ is the density of states per
spin (DOS) of the leads at the Fermi level (when the leads are
different $\rho_0^2=\rho_0^{\rm L}\rho_0^{\rm R}$ where $\rho_0^\alpha$
is the DOS in the lead $\alpha$).  The applied voltage is represented
by the difference in chemical potentials, $\mu_{\rm L} - \mu_{\rm R} =
V$. We also assume  $V\gg T_K\approx |T_{{\rm RL}}^{({\sf ex})}|
\exp(-|T_{{\rm RL}}^{({\sf ex})}|^{-2}\rho_0^{-2})$, and $T_0 \gg
T_{\alpha\beta}^{({\sf ex})}$ ($\alpha, \beta \in \{ {\rm R,  L }\} $).
We assume in the following that $B\rho_0\ll 1$ and $V\rho_0\ll 1$; in
other words, the lead-electron bandwidth $D\sim 1/\rho_0$ is the
largest energy scale. The leads are supposed to be in a voltage driven
thermal equilibrium at temperature $T$. ${\cal H}_{\sf ref}$ describes
spin-flip scattering of an electron from one lead back into the same
lead, and ${\cal H}_{\sf tr}$ represents both direct and spin-assisted
tunneling between leads. In the case of fully polarized leads, the
summation in the Hamiltonian is taken only over one type of electron
spin.

            In the model with spin-orbit interaction proposed by
Levitov and Rashba \cite{LevitovRashba} the spin precession induces
electron density oscillations inside the dot, and thus an electric
field inside and around the dot, ${\bf E}=({\bf n}\wedge({\bf
S}\wedge{\bf L}))$, where ${\bf L}$  is the orbital momentum, while
${\bf n}$ is a polar vector allowed by the symmetry of  the system.
This electric field modulates the tunneling barrier between the leads
and the dot. Hence, the tunneling matrix element acquires,  in addition
to the non-relativistic exchange term,  a dependence on the localized
spin ${\bf S}$ which is similar for both spin  components $\sigma$ of
the tunneling electrons.  To describe such relativistic corrections  we
need to add  to the transfer matrix elements in the Hamiltonian,
Eq.~(\ref{Ham}), the spin-orbit terms. The compound matrix elements are
finally given as
\begin{eqnarray}
(\hat{T}_{\alpha\alpha})_{\sigma\sigma'}&=&T_{\alpha\alpha}^{({\sf
ex})}{\bf S}\cdot\vec{\bm{\sigma}}_{\sigma\sigma'} +
T_{\alpha\alpha}^{({\sf so})}{\bf S}\cdot{\bf l} \ \delta_{\sigma\sigma'}, \\
(\hat{T}_{{\rm RL}})_{\sigma\sigma'}&=& (T_0+T_{{\rm RL}}^{({\sf
so})}{\bf S}\cdot{\bf l}) \
\delta _{\sigma\sigma'}+ T_{{\rm RL}}^{({\sf ex})}{\bf S} \cdot
\vec{\bm{\sigma}}_{\sigma\sigma'} .
\end{eqnarray}
Here the unit pseudo-vector ${\bf l}$ indicates what projections  of
the localized spin are modulating the tunneling matrix elements
depending upon the geometry of the system. We can estimate
$T_{\alpha\beta}^{({\sf ex})}\sim r^2E_0$, where $E_0$ is  an energy of
the order of an atomic energy, while $r$ is a small dimensionless
parameter which  describes the weak overlap of the lead-electron
wave-function to that of the molecule with the spin. For  spin-orbit
tunneling we have an additional small relativistic multiplicative
factor $\bm{\beta}^2=v^2/c^2$, i.e., $T_{\alpha\beta}^{({\sf so})}\sim
\bm{\beta}^2  T_{\alpha\beta}^{({\sf ex})}$,  where $v$ is a typical
electron velocity.

Finally, we presume that the spin also couples to the environment due
to the spin-phonon coupling, the coupling with nuclear spins, etc.  To
account for this spin relaxation mechanism we introduce the relaxation
rate $\Gamma_{{\sf env}}$.

In order to put our model in perspective let us compare it to
other models used in the literature. For the quantum dot studied in
Refs.~\cite{KorotkovAverin,RuskovKorotkov} the Hamiltonian for
spinless fermion tunneling is given by
\begin{eqnarray}
H&=&\sum_{\alpha, m}\!
\epsilon^{\;}_{m\alpha} c^{\dagger}_{\alpha m} c^{\;}_{\alpha m} +
\frac{b_z}{2} (d_{1} ^\dagger d_{1} - d_{2}^\dagger d_{2}) + \\
&& \frac{b_x}{2}(d^\dagger_{1}d^{\;}_{2}+d^\dagger_{2}d^{\;}_{1})+H_{\sf tr},
\nonumber\\
H_{\sf tr}&=&\sum_{mn}[ M + \frac{\Delta
M}{2} (d_{1} ^{\dagger} d_{1} - d_{2}^{\dagger} d_{2})]
\ c_{{\rm L}m}^\dagger c^{\;}_{{\rm R} n} +
{\rm H.c.}, \nonumber
\end{eqnarray}
where $d_{1,2}^\dagger$ ($d_{1,2}^{\;}$) creates (annihilates) an
electron in the state 1 or 2 of the dot. Here $b_z/2$ is the energy
splitting  between states 1 and 2, $b_x/2$ is the tunneling matrix
element connecting them, and tunneling of electrons between the leads
depends on the population of the states 1 and 2. The difference
between this and the TvS model (with $H_{\sf ref}$=0) is that the
couplings to the leads only depend upon the difference of fillings of
states 1 and 2 or, making correspondence with the spin operators, the
Hamiltonian ${\cal H}_s$ is
\begin{eqnarray}
{\cal H}_s&=&b_xS_x+b_zS_z, \label{transl} \\
S_x&=&(1/2)(d^\dagger_{1}d^{\;}_{2}+d^\dagger_{2}d^{\;}_{1}), \nonumber \\
S_z&=&(1/2)(d_{1} ^\dagger d_{1} - d_{2}^\dagger d_{2}),
\nonumber
\end{eqnarray}
while the coupling, ${\cal H}_{{\sf tr}}$ depends only on one spin
component, $S_z$ of the probed system, instead of all of them  as in
the exchange term ${\bf S} \cdot \vec{\bm{\sigma}}$. Moreover, the
leads electrons are fully polarized along the $z$-axis.

\section{Majorana-Keldysh formulation of tunneling via a single spin}
\label{3}

It is assumed that for sufficiently long times the composite
(spin-electrons) system reaches a dc (non-equilibrium) steady state
(which does not depend upon the initial conditions). We use the
Majorana fermion representation for the spin and the Keldysh
diagrammatic technique \cite{Keldysh} to describe it. This technique is
nowadays a standard method used in non-equilibrium phenomena and its
basic description can be found in standard textbooks such as Ref.
\cite{Mahan}. Its application to Kondo-like problems is described in
Ref. \cite{ParcolletHooley}.

            Since spin operators are not amenable to the application of
the Wick's theorem, and hence are not appropriate for diagrammatic
techniques, we represent spin-$1/2$ operators $S_\mu$ in terms of three
Majorana fermions $\eta_\mu$, where $\mu \in \{x,y,z\}$, satisfying the
following mapping and relations
\begin{eqnarray}
&&S_x=- i \eta_y\eta_z, \ \ \ S_y=-i\eta_z\eta_x, \ \ \
S_z=-i\eta_x\eta_y,            \label{Majorana} \\
&& (\eta_\mu)^\dagger=\eta_\mu, \ \ \ \{ \eta_\mu, \eta_\nu \}=
\delta_{\mu\nu} .
\label{relation}
\end{eqnarray}
The Hilbert space for Majorana fermions is eight-dimensional (8$d$) and
the transformation from the spin 1/2 to the Majorana fermion
representation may be performed in the following way. In general, the
spin expectation values that we need to calculate are of the form of a
trace, ${\rm Tr}[F\{S_\mu\}]$, over the two-dimensional (2$d$) spin
Hilbert space. Here $F$ is some functional of the spin operators. We
replace the spin operators by Majorana fermion operators using
Eq.~(\ref{Majorana}) and we represent the Majorana fermion operators as
\begin{equation}
\eta_\mu=(c_\mu^\dagger+c^{\;}_\mu)/\sqrt{2},
\end{equation}
where $c^\dagger_\mu$ and $c^{\;}_\mu$ are the creation and
annihilation operators of a type $\mu \in \{ x,y,z \}$ fermion
satisfying the canonical anti-commutation relation $\{c^{\;}_\mu,
c_\nu^\dagger\}=\delta_{\mu\nu}$. This representation guarantees that
Eq.~(\ref{relation}) for the Majorana fermion operators are satisfied.
A possible orthonormal basis for the 8$d$ Hilbert space ${\cal H}_M$ is
$\varphi_1=\Phi_0$, where $\Phi_0$ is the vacuum,
$\varphi_2=c_x^\dagger\Phi_0$, $\varphi_3=c_y^\dagger\Phi_0$,
$\varphi_4=c_z^\dagger\Phi_0$, $\varphi_5=c_x^\dagger
c_y^\dagger\Phi_0$, $\varphi_6=c_y^\dagger c_z^\dagger\Phi_0$,
$\varphi_7=c_z^\dagger c_x^\dagger\Phi_0$, and $\varphi_8=c_x^\dagger
c_y^\dagger c_z^\dagger \Phi_0$. ${\cal H}_M$ may be written as a
direct sum of four orthogonal 2$d$ subspaces, ${\cal H}_M={\cal H}_{1+}
\oplus {\cal H}_{1-}\oplus {\cal H}_{2+}\oplus {\cal H}_{2-}$, which
are invariant under the action of the spin-1/2 operators $S_\mu$.
Here, basis sets for ${\cal H}_{1+}$, ${\cal H}_{1-}$, ${\cal H}_{2+}$,
and ${\cal H}_{2-}$ are: $\{(\varphi_6+
i\varphi_7)/\sqrt{2},(\varphi_1+i\varphi_5)/\sqrt{2}\}$,
$\{(\varphi_1-i\varphi_5)/\sqrt{2}, (\varphi_6-
i\varphi_7)/\sqrt{2}\}$, $\{(\varphi_2+ i\varphi_3)/\sqrt{2},
(\varphi_4+ i\varphi_8)/\sqrt{2}\}$, and $\{ (\varphi_4-
i\varphi_8)/\sqrt{2},(\varphi_2- i\varphi_3)/\sqrt{2}\}$, respectively.
In other words, the  Majorana fermion representation of the spin-$1/2$
operator  $S_{\mu}$ is an irreducible one in any of the four mentioned
subspaces. Furthermore, these representations are equivalent to each
other. Therefore, we can write down
\begin{equation}
{\rm Tr}[F\{S_\mu\}]=\frac{1}{4}{\rm Tr}[\tilde{F}\{\eta_\mu\}],
\end{equation}
where the second trace is computed in the extended Hilbert
space ${\cal H}_M$. (The trace is invariant under a change of basis in
${\cal H}_M$). Now one can prove Wick's theorem in the same way as it
was proved for the free Fermi and Bose fields.

The time-ordered averages in the Keldysh technique are taken with the
help of the evolution operator $S_{\sf C}$ along a closed time contour
(its part running from $- \infty$ to $+ \infty$ denoted by $+$ and the
part from $+ \infty$  to $- \infty$ by $-$). One defines the four
(non-independent) real-time Green's functions
\begin{eqnarray}
G^{++}_{\psi}(t,t')&=&-i \langle T \psi(t)\psi ^\dagger (t') \rangle,
\nonumber \\
G^{-+}_{\psi}(t,t') &=&-i \langle \psi(t)\psi ^\dagger (t') \rangle,
\nonumber \\
G^{+-}_{\psi}(t,t') &=& i \langle \psi ^\dagger(t')\psi (t) \rangle ,
\nonumber \\
G^{--}_{\psi}(t,t')&=&-i \langle \widetilde{T} \psi(t)\psi ^\dagger
(t') \rangle ,
\label{Gpsi}
\end{eqnarray}
which can be compactly written as a matrix $G_{\psi}(t,t')=-i
\greenf{\psi(t)\psi ^\dagger (t')}$. Here $T$ and $\widetilde{T}$ are
the time- and anti-time-ordering operators on the $+$ and $-$ parts of
the Keldysh contour and $\psi$ represents an arbitrary fermionic field.
They are related through:
$G^{++}_{\psi}+G^{--}_{\psi}=G^{+-}_{\psi}+G^{-+}_{\psi}$, and the
brackets $\rangle$ have the meaning of either a pure state or a
distribution of the available phase space of the interacting system.

The four Green's functions (the ``$\pm$ basis'') can be expressed in
terms of the retarded, advanced and Keldysh Green's functions
(``Larkin-Ovchinnikov (LO) basis'')
\begin{eqnarray}
G_{\psi}^R(t,t')&=&-i\theta (t-t') \langle \{\psi (t),\psi ^\dagger
(t') \}  \rangle ,  \nonumber \\
G_{\psi}^A(t,t')&=&i \theta(t'-t) \langle \{\psi (t),\psi ^\dagger (t')
\} \rangle , \nonumber  \\
G_{\psi}^K(t,t')&=&-i \langle  [\psi (t),\psi^\dagger (t') ] \rangle ,
\label{Kel}
\end{eqnarray}
by use of the transformation
\begin{equation}
\left( \begin{array}{cc}
G_{\psi}^R & G_{\psi}^K\\
0 & G_{\psi}^A
\end{array}
\right)= \frac{1}{2}
\left( \begin{array}{cc}
1 & 1 \\
1 & -1
\end{array}
\right) \left( \begin{array}{cc}
G_{\psi}^{++} & G_{\psi}^{+-}  \\
G_{\psi}^{-+} & G_{\psi}^{--}
\end{array}
\right)
\left( \begin{array}{cc}
1 & 1 \\
-1 & 1
\end{array}
\right).
\nonumber
\end{equation}
In general, the retarded and advanced functions are Hermitian
conjugates of each other, i.e.,
$G_{\psi}^R(t,t')=(G_{\psi}^A(t',t))^*$. Furthermore, in thermal
equilibrium the fluctuation-dissipation theorem connects them through
\begin{equation}
G_{\psi}^K(\omega)= h_{\sf eq} (\omega)
[G_{\psi}^A(\omega)-G_{\psi}^R(\omega)] ,
\end{equation}
with the thermal distribution function of fermions $h_{\sf eq}(\omega)=
-\tanh(\omega/2T)$. As in Ref.~\cite{ParcolletHooley} we assume that
for sufficiently long times the spin reaches a dc (non-equilibrium)
steady state which does not depend on the initial conditions.  In such
dc steady state the correlation functions depend upon the time
difference and the new distribution function $h(\omega)$ differs from
$h_{\sf eq}(\omega)$. It should be determined as a stationary solution
of the quantum kinetic equation.

Firstly, we introduce the bare Green functions for electrons in the
presence of an applied magnetic field (or internal exchange field in
the case of magnetically ordered leads) ${\bf B}_{\alpha}$ which
induces a spin polarization of electrons in the directions ${\bf
m}_{{\alpha}0}={\bf B}_{\alpha}/B_{\alpha}$.
Since the leads are assumed to be coupled to a thermal
bath, the bare lead-electron Green's functions are in a voltage driven
thermal equilibrium. They have the matrix form in the space of electron
spin ($[{\bf G}_{\alpha}(t,t')]_{\sigma\sigma'}
=-i\greenf{c^{\;}_{\alpha\sigma}(t) c^{\dagger}_{\alpha\sigma'}(t')}$,
where $c^{\;}_{\alpha\sigma} = \sum_n c^{\;}_{\alpha n\sigma}$)
\begin{eqnarray}
{\bf G}_{\alpha}(\omega)&=&G_{\alpha 1}(\omega) \ \one+G_{\alpha 2}
(\omega) \ {\bf m}_{\alpha0}\cdot \vec{\bm{\sigma}}, \\
G_{\alpha 1,2}(\omega)&=&\frac{1}{2}
[G_{\alpha+}(\omega)\pm G_{\alpha-}(\omega)],  \nonumber\\
G_{0 \alpha\kappa}^R(\omega)&=&\int d \epsilon  \
\frac{\rho_{\alpha\kappa}(\epsilon)}{\omega - \epsilon + i0^+},  \nonumber\\
G_{0 \alpha\kappa}^K (\omega) &=& 2 \pi i \ h_{0 \alpha}(\omega) \
\rho_{\alpha\kappa}(\omega),  \nonumber\\
h_{0\alpha} (\omega) &=&  h_{\sf eq}(\omega- \mu
_{\alpha})=-\tanh[\frac{\omega-\mu_{\alpha}}{2T}], \nonumber \\
\rho_{\alpha\kappa}(\epsilon)&=&\rho(\epsilon-\kappa
B_\alpha-\mu_{\alpha}), \nonumber
\end{eqnarray}
where $\kappa=\pm$ and $\rho(\omega)$ is the bare electron density of
states (per  spin) which we assume to be similar in both leads. We
consider here only local Green's functions, so the eigenstate index $n$
can be dropped. Note, that for fully polarized electrons
($B_{\alpha}\gg D$) we obtain $G_{\alpha 1}(\omega)=G_{\alpha
2}(\omega)$, while for weakly  spin polarized electrons ($B_{\alpha}\ll
D$) $G_{\alpha 2}(\omega)\ll G_{\alpha 1}(\omega)$.

The propagators for the Majorana fields can be compactly written as
$G_{\mu\nu}(t,t')=-i\greenf{\eta_\mu(t)\eta_\nu(t')}$.  The bare
propagators satisfy $G_{xx}^0=G_{yy}^0$, $G_{yx}^0=-G_{xy}^0$,
$G_{xz}^0=G_{zx}^0=G_{yz}^0=G_{zy}^0=0$, due to the symmetry under
rotations about the $z$-axis. A useful property holds for the Keldysh
components of the spin Green's functions: Since Majorana fermion
operators are Hermitian, from the properties of the commutator, it
turns out that $G_{\mu\nu}^K(t,t')=-G_{\nu\mu}^K(t',t)$. In the
frequency $\omega$ representation, for time-translation invariant
solutions, this means that $G_{\mu\nu}^K(\omega)= -
G_{\nu\mu}^K(-\omega)$, and therefore $G_{\mu\mu}^K(\omega)$  is an odd
function of $\omega$.

To write down Dyson's equation, it is convenient to use a basis in
which the bare propagator is diagonal. Defining the canonical fermion
operators $f=(\eta_x - i \eta_y)/\sqrt{2}$ and $f^{\dagger}=(\eta_x+i
\eta_y)/\sqrt{2}$, with the associated Green's functions
$G_{ff^{\dagger}}(t)=-i\greenf{f(t)f^\dagger(0)}$ and
$G_{f^{\dagger}f}(t)=-i\greenf{f^\dagger(t)f(0)}$, one has the
following expressions for the bare propagators in the
$(f,f^{\dagger},\eta_z)$ basis in the steady state in the lowest order
of perturbation theory
\begin{eqnarray}
\!\!\!\!G_{zz}^{0 R}(\omega)\!\!&=&\!\!\frac{1}{\omega +is},\ \ \ \ \ \
\ \ \   G_{zz}^{0 K}(\omega)= \frac{2 i h_z(\omega) s}{\omega ^2
+s^2} , \label{free}\\
\!\!\!\!\!\!\!\!\!\!\!\!G_{ff^{\dagger}}^{0 R}(\omega)\!\!&=&\!\!\frac{1}{\omega
-B+is}, \ G_{ff^{\dagger}}^{0 K}(\omega) = \frac{2 i h_f(\omega)s}{(\omega-B)^2 +s^2} ,
\nonumber \\
\!\!\!\!\!\!\!\!\!\!\!\!G_{f^{\dagger}f}^{0 R}(\omega)\!\!&=&\!\!\frac{1}{\omega
+B+is}, \ G_{f^{\dagger}f}^{0 K}(\omega) = \frac{2 i h_{f^{\dagger}}(\omega)s}
{(\omega+B)^2 +s^2} , \nonumber
\end{eqnarray}
where $s\rightarrow 0^+$ is a small regulator (a width due to an
infinitesimally small coupling to a thermal bath), and $B$ is the total
effective magnetic field acting on the spin (see next section). In the
thermal equilibrium state
$h_z(\omega)=h_f(\omega)=h_{f^{\dagger}}(\omega)=-\tanh(\omega/2T)$.

\begin{figure}
\centerline{\epsfxsize=7cm \epsfbox{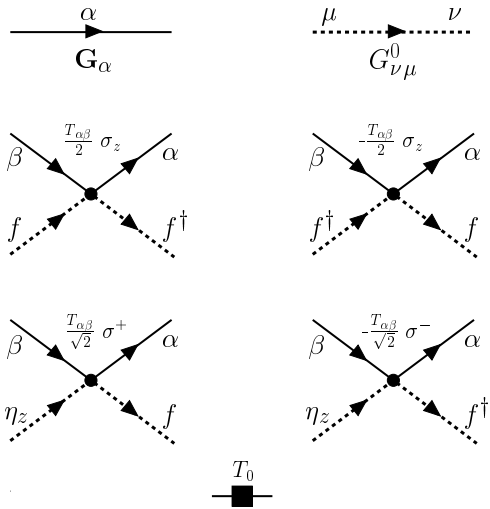}}
\caption{Diagrams for: -the lead electron Green's function (solid
line); -the spin Green's function; -vertices for the exchange
interaction.}
\label{fig2}
\end{figure}

The perturbative calculations of various physical quantities are very
conveniently performed by using the matrix form of the Green's
functions in the Keldysh space. Graphical representations can then be
done by Feynman diagrams with lines representing matrices of Green's
functions, where at each vertex of an internal point is assigned an
additional $+$ or $-$ factor (due to the opposite direction of time
integration for the points on the $-$ part of the Schwinger-Keldysh
contour). The interaction vertices of Majorana fermions with electrons
follow from the form of the tunneling Hamiltonian (\ref{Ham})
\begin{eqnarray}
T_{\alpha\beta}^{({\sf ex})}{\bf S}\cdot\vec{\bm{\sigma}}\!\!&=&\!\!
\frac{T_{\alpha\beta}^{({\sf
ex})}}{2}\sigma_z(f^{\dagger}f\!-\!ff^{\dagger})+
\frac{T_{\alpha\beta}^{({\sf ex})}}{\sqrt{2}}(
f\sigma^+\!-\!f^{\dagger}\sigma^-)\eta_z,   \nonumber           \\
T_{\alpha\beta}^{({\sf so})}{\bf S}\cdot\vec{\bm{\sigma}}\!\!&=&\!\!
\frac{T_{\alpha\beta}^{({\sf
so})}}{2}l_z(f^{\dagger}f\!-\!ff^{\dagger})+
\frac{T_{\alpha\beta}^{({\sf so})}}{\sqrt{2}}(
f \ l^+\!-\!f^{\dagger} \ l^-)\eta_z,      \nonumber        \\
\sigma^{\pm}&=& \sigma_x\pm i\sigma_y, \ \   l^{\pm}= l_x\pm il_y ,
\end{eqnarray}
accounting for the relation $ff^{\dagger}+f^{\dagger}f=1$.
We represent the electron and spin Green's functions as well
as the vertices in Fig.~\ref{fig2}.

The interacting Green's function for the spin, ${\bf G}$, satisfies the
Dyson's equation
\begin{equation}\label{Dyson}
{\bf G}^{-1}(\omega)={\bf G}_0^{-1}(\omega)-{\bf \Sigma}(\omega) ,
\end{equation}
for the $6\times6$ matrices (in the tensor product of the $x,y,z$ space
and the Keldysh space), where the free propagator is given by
Eqs.~(\ref{free}). To lowest order, the imaginary part of the
self-energy ${\bf \Sigma}(\omega)$ is  of order
$T_{\alpha\beta}^2\rho_0^2$ (its $\omega$-dependent real part induces a
trivial small energy shift that we neglect in the following).
Calculating ${\bf G}(\omega)$ and neglecting  terms of order
$T_{\alpha\beta}^4\rho_0^4$  we can take the $2\times 2$ blocks
$\Sigma_{ff}$, $\Sigma_{f^{\dagger}f^{\dagger}}$, $\Sigma_{fz}$,
$\Sigma_{f^{\dagger}z}$, $\Sigma_{zf}$ and $\Sigma_{zf^{\dagger}}$ as
zeroes. Then, ${\bf G}^{-1}(\omega)$ has nonvanishing off-diagonal
blocks $ff^{\dagger}$ and $f^{\dagger}f$,
\begin{equation}\label{block}
{\bf G}_{ff^{\dagger}}^{-1}= \left(
\begin{array}{cc}
\omega - B - \Sigma _{ff^{\dagger}} ^R (\omega)   &
-\Sigma _{ff^{\dagger}} ^K (\omega)   \\
0 & \omega - B - \Sigma _{ff^{\dagger}} ^A (\omega)
\end{array}  \right),
\end{equation}
and the diagonal block $zz$. Inverting such block matrix ${\bf
G}^{-1}(\omega)$, we obtain the matrix  ${\bf G}(\omega)$ with a
similar block structure, where the block $ff^{\dagger}$ is given as
\begin{equation}
\label{DysonB}
{\bf G}_{ff^{\dagger}}= \left(
\begin{array}{cc}
\frac{1}{\omega - B - \Sigma _{f^{\dagger}f} ^R (\omega)}  &
\frac{\Sigma _{f^{\dagger}f} ^K (\omega)}
{|\omega - B - \Sigma _{f^{\dagger}f} ^R (\omega) |^2}  \\
0 & \frac{1}{\omega - B - \Sigma _{f^{\dagger}f} ^A (\omega)}
\end{array}  \right),
\end{equation}
while for ${\bf G}_{f^{\dagger}f}$ we replace $ f\rightarrow
f^{\dagger}$ and $B\rightarrow -B$. For $G_{zz}$ we replace $zz$ for
${f^{\dagger}f}$ and put $B=0$.

Parcollet and Hooley \cite{ParcolletHooley} have found that if one
starts from the equilibrium distribution functions for Majorana
fermions, the perturbation theory breaks down when one takes the limit
of zero coupling to the leads prior to taking the limit of zero
coupling to the thermal bath. The reason for this non-commutativity of
the limits is that in a  non-equilibrium steady state the distribution
function of the system can deviate from the equilibrium one
significantly in the long-time limit even for very weak tunneling.
Hence, at long times it cannot be calculated perturbatively.
One can, nevertheless, use the perturbation theory built upon the
appropriate bare Green's functions with correct zeroth-order
distribution functions $h_z(\omega)$, and $h_{f,f^{\dagger}}(\omega)$
which are stable with respect to weak perturbations.  These
distributions should be obtained self-consistently, i.e., $h_z(\omega)$
and $h_{f,f^{\dagger}}(\omega)$ computed using second order
perturbation theory have to be the same as the zero order distribution
function.  Hence, assuming that after a long time the system is in a
steady state, we can define a distribution function $h_f(\omega)$
obeying the self-consistency equations
\begin{eqnarray}
G_{ff^{\dagger}}^K(\omega)&=& h_f (\omega) [G_{ff^{\dagger}}^A (\omega)
-G_{ff^{\dagger}}^R (\omega)],   \nonumber\\
h_f(\omega)&=& \frac{\Sigma ^K_{f^{\dagger}f}(\omega)}
{\Sigma^A _{f^{\dagger}f}(\omega)-\Sigma^R_{f^{\dagger}f}(\omega)},
\label{hB}
\end{eqnarray}
where $\Sigma_{f^{\dagger}f}(\omega)$ is calculated in the second order
perturbation  theory with respect to the electron-spin interaction.
$h_{f^{\dagger}}(\omega)$ and $h_z(\omega)$ have to be obtained in a
similar way. Therefore, one needs to take the zeroth-order  Green's
functions for the spin in the form of Eq.~(\ref{free}), calculate the
second order self-energies $\Sigma ^K_{f^{\dagger}f}(\omega)$ and
$\Sigma^A_{f^{\dagger}f}(\omega)- \Sigma ^R_{f^{\dagger}f} (\omega)$,
and solve Eq.~(\ref{hB}). This procedure, as well as taking the limits
$s \rightarrow 0^+$ and $T_{\alpha\beta} \rightarrow  0$, was carefully
outlined by Parcollet and Hooley \cite{ParcolletHooley}.

A useful relation between the distribution functions $h_f(\omega)$ and
$h_{f^\dagger}(\omega)$ can be derived. Using the definitions in
Eq.~(\ref{Kel}) and the commutator and anticommutator properties one
can show that
\begin{eqnarray}
G_{f^{\dagger}f}^K (-\omega) &=& -G_{ff^{\dagger}} ^K (\omega), \\
G_{f^{\dagger}f}^A (-\omega) - G_{f^{\dagger}f}^R (-\omega) &=&
G_{ff^{\dagger}}^A (\omega) -G_{ff^{\dagger}}^R (\omega) , \nonumber
\end{eqnarray}
and using the definition of Eq.~(\ref{hB}) one obtains that
\begin{equation}
h_f (\omega) =-h_{f^{\dagger}}(-\omega), \ \ h_{z}(\omega)=-h_{z}(-\omega),
\end{equation}
so that $h_{z}(0)=0$.

\section{Effect of the tunneling electrons on the spin dynamics}
\label{sec4}

\subsection{Effective dc magnetic field acting on the spin}
\label{sec4a}
\begin{figure}
\centerline{\epsfxsize=8cm \epsfbox{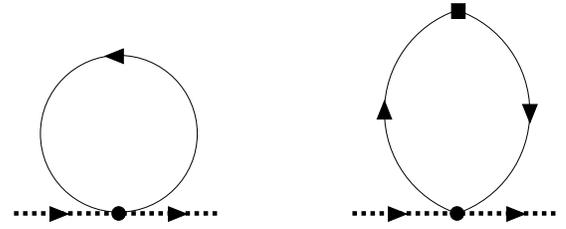}}
\caption{Diagrams for the self-energy due to
the electron-spin interaction that contribute to the effective dc
magnetic field. }
\label{fig3}
\end{figure}
The diagrams shown in Fig.~\ref{fig3} lead to a renormalization of the
effective dc magnetic field acting on the spin when the electrons are
polarized. To first order in the electron-spin interaction (the first
diagram in Fig.~\ref{fig3}) the  additional dc magnetic field caused by
the tunneling electrons is given by
\begin{eqnarray}
\mu {\bf B}_{T}&=&2{\rm Re}
\sum_{\alpha,n, \sigma, \sigma'}
T_{\alpha\alpha}^{({\sf ex})}\langle c^{\dagger}_{\alpha n \sigma}
\vec{\bm{\sigma}}_{\sigma\sigma'} c^{\;}_{\alpha n
\sigma' }\rangle \nonumber \\
&=&4 \sum_\alpha s_{\alpha}{\bf m}_{\alpha 0}T_{\alpha\alpha}^{({\sf
ex})} , \label{bren}
\end{eqnarray}
neglecting spin-orbit corrections. Here $s_{\alpha}$ is the degree of
the electron  spin polarization in the lead $\alpha$, i.e., the ratio
of the electron magnetization to the optimal magnetization. This
additional magnetic field  originates due to the exchange interaction
between the spin and the electrons in the leads, and   it is voltage
independent.

We may estimate this contribution using the expression for the
tunneling resistance $R_T=\hbar/(4\pi e^2T_0^2\rho_0^2)$. For $R_T$
=100 Mohm we obtain  $T_0\rho_0\approx 1.6\cdot 10^{-3}$ and taking
$T_{\alpha\alpha}^{({\sf ex})}\approx 0.1T_0$ we  estimate at
$\rho_0\approx $ eV$^{-1}$ the additional magnetic field $B_T\approx
16s_{\alpha}$ T. When electrons are polarized by the magnetic (or
exchange) field $B_{\alpha}$ such that $B_{\alpha}\rho_0\ll 1$, we
estimate $B_T\approx 10^{-3}B_{\alpha}$. The second diagram in
Fig.~\ref{fig3} is smaller by a  factor $T_0\rho_0$ and, thus, may be
neglected.
\begin{figure}
\centerline{\epsfxsize=8cm \epsfbox{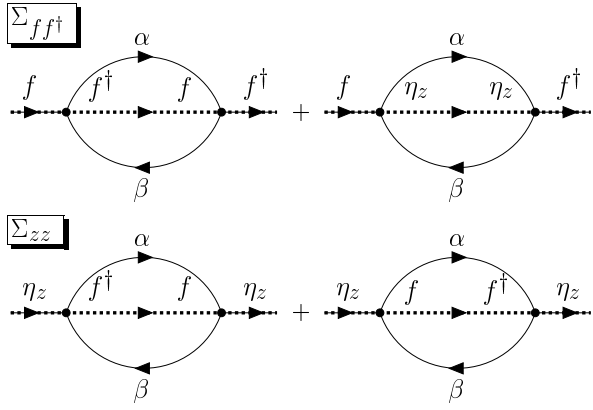}}
\caption{Diagrams for the self-energy to second order in $T_{\alpha\beta}$.}
\label{fig4}
\end{figure}

In the following we choose the $z$-axis along the total field ${\bf
B}={\bf B}_0+{\bf B}_T$. By  ${\bf m}_{\alpha}$ we denote the direction
of electron spin polarization in this new coordinate system. In the
case that  ${\bf m}_{{\rm R}0}={\bf m}_{{\rm L}0}={\bf m}_0$, and ${\bf
m}_0$ has  the components $m_{0x},m_{0z}$ in the coordinate system with
the $z$-axis  aligned along the bare field ${\bf B}_0$, we get that
${\bf m}$ has  coordinates $m_x,m_z$ in the coordinate system where the
$z$-axis is aligned  along ${\bf B}$ with
\begin{eqnarray}
m_x&=&\frac{\sqrt{B_0^2-({\bf B}_0\cdot {\bf m}_0)^2}}
{\sqrt{B_0^2+B_T^2+2B_T{\bf B}_0\cdot {\bf m}_0}},   \\
m_z&=&\frac{({\bf B}_0\cdot{\bf m}_0)+B_T}
{\sqrt{B_0^2+B_T^2+2B_T{\bf B}_0\cdot {\bf m}_0}}.
 \nonumber
\end{eqnarray}

\subsection{Self-energy of Majorana fermions and steady state
distribution function}
\label{sec4b}

The diagrams shown in Figs.~\ref{fig4} determine the imaginary part of
the Majorana fermion self-energies which depend upon $T$, $V$, ${\bf
m}_{\alpha}$ and ${\bf B}$.  We start by considering only the
interaction of the spin with the tunneling electrons, assuming 
$\Gamma_{{\sf env}}=0$. In this case, the spin is precessing freely
between consecutive passings of tunneling electrons which change the
phase of precession randomly, preserving time-translation invariance on
the average in the steady state. In the quasi-classical approach
\cite{BulaevskiiOrtiz} these passings are replaced  by an effective
classical magnetic field representing white noise. On the other hand,
in the quantum description the electrons are treated as tunneling
particles with initial energy  $\epsilon_n+V$ and final energy
$\epsilon_{n'}$. The tunneling process now has similarities with the
scattering of an individual particle by the spin. The difference is
that both the initial and final energy of the tunneling electrons are
not fixed, in our model being restricted only by the bands of the
leads-electrons.

The diagrams shown in Fig.~\ref{fig4} lead to the self-energy
\begin{eqnarray}
\Sigma_{ff^{\dagger}}^{+-}(\omega)&=&-\sum_{\alpha\beta}T_{\alpha\beta}^2
\int\frac{dudvd\epsilon}
{(4\pi)^2}\delta(\omega+v-u-\epsilon)\times \nonumber \\
&&\{
4{\rm Tr}[\sigma_z{\bf G}_{\alpha}^{+-}(u)\sigma_z{\bf G}_{\beta}^{-+}(v)]
G_{f^{\dagger }f}^{+-}(\epsilon)+ \nonumber\\
&&2{\rm Tr}[\sigma^+{\bf G}_{\alpha}^{+-}(u)
\sigma^-{\bf G}_{\beta}^{-+}(v)]
G_{zz}^{+-}(\epsilon)\} .
\end{eqnarray}
The expression for $\Sigma_{ff^{\dagger}}^{-+}$ can be similarly
obtained. Note, that $\Sigma_{ff^{\dagger}}^{+-}$ depends on
$h_{f^{\dagger}}(-B)$ via $G_{f^{\dagger}f}^{+-}$, while
$\Sigma_{f^{\dagger}f}^{+-}$ depends on $h_{f}(B)$. Then,
Eq.~(\ref{hB}) is a closed equation for $h_f(\omega)$. The other
diagrams in Fig.~\ref{fig4} give
\begin{eqnarray}
\Sigma_{zz}^{+-}(\omega)&=&-\sum_{\alpha\beta}T_{\alpha\beta}^2
\int\frac{dudvd\epsilon}
{8\pi^2}\delta(\omega+v-u-\epsilon)\times \nonumber \\
&&\{
{\rm Tr}[\sigma^-{\bf G}_{\alpha}^{+-}(u)\sigma^+{\bf G}_{\beta}^{-+}(v)]
G_{f^{\dagger}f}^{+-}(\epsilon)+ \nonumber\\
&&{\rm Tr}[\sigma^+{\bf G}_{\alpha}^{+-}(u)
\sigma^-{\bf G}_{\beta}^{-+}(v)]
G_{ff^{\dagger}}^{+-}(\epsilon)\}.
\end{eqnarray}
The traces for the spin matrices in the expressions for the self-energy
can be performed with the help of the relations
\begin{eqnarray}
\frac{1}{2}\sum_{\gamma\delta}m_{\alpha \gamma}m_{\beta \delta}{\rm
Tr}[\sigma_\mu\sigma_\gamma\sigma_\nu\sigma_\delta]&=&
\nonumber \\
&&\hspace*{-3.5cm}-{\bf m}_{\alpha}\cdot{\bf m}_{\beta} \ \delta_{\mu\nu}+m_{\alpha
\mu}m_{\beta \nu}+m_{\alpha \nu}
m_{\beta \mu}, \nonumber\\
\sum_\nu m_{\alpha \nu}{\rm
Tr}[\sigma_\mu\sigma_\nu]=2m_{\alpha \mu}, \ && \!\!
{\rm Tr}[\sigma_x\sigma_y\sigma_z]=2i.
\end{eqnarray}
The transformation from the "$\pm$ basis" to the ``LO basis'' for the
self-energy is the same as for the inverse Green's function
\begin{equation}
\left( \begin{array}{cc}
\Sigma_{\psi}^R & \Sigma_{\psi}^K\\
0 & \Sigma_{\psi}^A
\end{array}
\right)= \frac{1}{2}
\left( \begin{array}{cc}
1 & -1 \\
1 & 1
\end{array}
\right) \left( \begin{array}{cc}
\Sigma_{\psi}^{++} & \Sigma_{\psi}^{+-}  \\
\Sigma_{\psi}^{-+} & \Sigma_{\psi}^{--}
\end{array}
\right)
\left( \begin{array}{cc}
1 & 1 \\
1 & -1
\end{array}
\right),
\label{SLOto+-}
\nonumber
\end{equation}
which implies that $\Sigma_{\psi}^{++} + \Sigma_{\psi}^{+-} +
\Sigma_{\psi}^{-+} + \Sigma_{\psi}^{--}=0$ and, therefore,
$\Sigma^K=-(\Sigma^{+-}+\Sigma^{-+})$ and $\Sigma^{A}-\Sigma^{R}=
\Sigma^{-+}-\Sigma^{+-}$.

In this way, for the fully spin-polarized case,
$m_{\alpha}=m_{\beta}=1$, we obtain
\begin{eqnarray}
\Sigma_{ff^{\dagger},f^{\dagger}f}^K(\omega)\!\!\!&=&\!\!\!i\frac{\pi}{2}\rho_0^2
\sum_{\alpha,\beta}\!T_{\alpha \beta}^2 \{(1+ 2m_{\alpha z} m_{\beta
z}\!\! -\!{\bf m}_{\alpha} \!\cdot \!{\bf m}_{\beta}) \nonumber\\
&&  \hspace*{-2cm}[(\mu _{\alpha}-\mu _{\beta }-\omega \mp
B) \mp h_{f}(B)\
T\phi(\frac{\mu _{\beta } -\mu _{\alpha} +\omega \pm
B}{T} )]
\nonumber \\
&& \hspace*{-2.3cm}-(1- m_{\alpha z} m_{\beta z} \pm  m_{\alpha z} \mp  m_{\beta z})
(\mu _{\beta} -\mu _{\alpha} +\omega) \}
\end{eqnarray}
and
\begin{eqnarray}
[\Sigma^A-\Sigma^R]_{ff^{\dagger},f^{\dagger}f}(\omega)\!\!\!&=&\!\!\!
i\frac{\pi}{2}\rho_0^2\sum _{\alpha, \beta}T_{\alpha
\beta}^2 \times \\
&& \hspace*{-3.9cm} \{(1+
2m_{\alpha z} m_{\beta z} -{\bf m}_{\alpha} \cdot {\bf
m}_{\beta}) \nonumber\\
&& \hspace*{-3.9cm} [T\phi(\frac{\mu_{\beta}-\mu _{\alpha}+\omega
\pm B}{T})
\pm h_{f}(B)(\mu _{\beta } -\mu _{\alpha} +\omega \pm
B)]
\nonumber\\
&&\hspace*{-3.9cm} + (1- m_{\alpha z} m_{\beta z} \pm  m_{\alpha z} \mp m_{\beta z})
T\phi (\frac{\mu _{\beta} -\mu _{\alpha} +\omega}{T} ) \},
\nonumber
\end{eqnarray}
with the property
\begin{eqnarray}
{\rm Im}\Sigma_{ff^{\dagger},f^{\dagger}f}^A=-{\rm
Im}\Sigma_{ff^{\dagger},f^{\dagger}f}^R=\frac{1}{2}
(\Sigma^A-\Sigma^R)_{ff^{\dagger},f^{\dagger}f}. \nonumber
\end{eqnarray}
\begin{figure}
\centerline{\epsfxsize=9cm \epsfbox{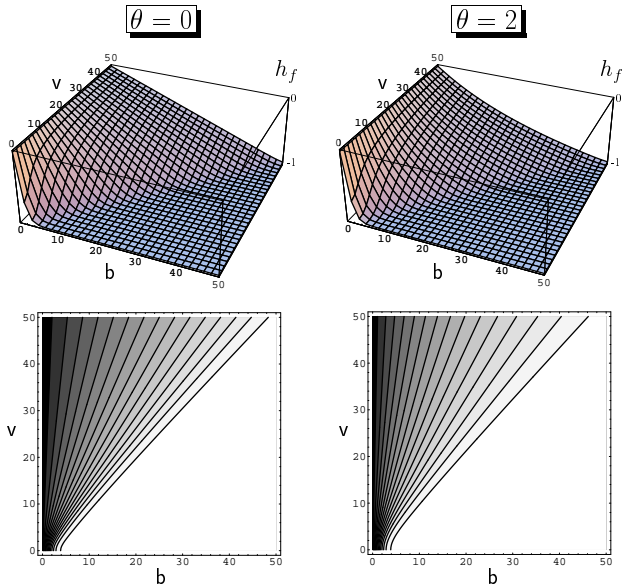}}
\caption{Upper panel displays the steady state distribution function
$h_f(B)$ in the situation $m_{{\rm R}z}=m_{{\rm L}z}=0$ as a function
of ${\sf b}$ and ${\sf v}$, for two different values of $\theta$. Lower
panel shows contour plots of the same function. The case ${\sf v}=0$
corresponds to the equilibrium distribution $h_{\sf eq}(B)$.}
\label{fig5}
\end{figure}
Assuming that $D\gg V,B,T$, we calculated the integrals over electron
energies,
\begin{equation}
\int du  \ G_{\alpha 1}^{+-}(u) G_{\beta 1}^{-+}(u+w),
\nonumber
\end{equation}
in the limit of large electron bandwidth, $D\rightarrow \infty$, and
using the relation $1 \pm h_{\sf eq} (\omega)=2n_F(\pm \omega)$  (where
$n_F(x)=[\exp(x)+1]^{-1}$ is the Fermi distribution function) the
resulting function $\phi(x)$ is
\begin{equation}
\phi(x)= \frac{x}{\tanh (x/2)}   \ .
\end{equation}

In the weak polarization limit, $B_{\alpha}\ll D$, there are
corrections to these expressions due to terms consisting of integrals
of the form
\begin{equation}
\int du \ G_{\alpha 2}^{+-}(u) G_{\beta 2}^{-+}(u+w) .
\nonumber
\end{equation}
To estimate them we assume that near the Fermi energy the electron
density of states may be approximated as
$\rho(\epsilon)\approx\rho_0(1+\epsilon/D)$. Then, these integrals are
proportional to the small parameter $(B_{\alpha}/D)^2$  and they vanish
in the limit $D\rightarrow \infty$. We obtain the expressions for
self-energies in the weak polarization limit from the fully polarized
results by putting ${\bf m}_{\alpha}=0$ and  multiplying the
resulting expressions by a factor 4 that accounts for both spin
projections of the electron.

In the next section we show that the imaginary part of the self-energy 
$\Sigma_{ff^{\dagger}}(\omega)$, $\Gamma_{\perp}={\rm
Im}\Sigma_{ff^{\dagger}}(\omega)$, describes dephasing of the spin
precession caused by the tunneling electrons. It determines the
current noise power as shown later. 

Therefore, at frequency $\omega = \pm B$, Eq.~(\ref{hB}) gives us a
self-consistent equation for $h_{f}(B)=-h_{f^{\dagger}}(-B)$ in the
case of full spin polarization.  This equation has a unique solution
for the distribution function  (${\sf b}=B/T$, ${\sf v}=V/T$)
\begin{eqnarray}
h_{f}(B)\!\!\!&=& \!\!\!-\frac{2{\sf b}(1-m_{{\rm R} z} m_{{\rm L} z}) \!-\!
2 {\sf v} ( m_{{\rm R} z} - m_{{\rm L} z})
\! + \!{\sf b} \ \theta}{\phi^+  (1-m_{{\rm R} z}
m_{{\rm L} z})\!-\!  \phi^-
(m_{{\rm R} z} - m_{{\rm L} z})\! + \!\phi ({\sf b}) \ \theta} ,
\nonumber \\
\theta &=& \frac{T_{{\rm RR}}^2 ( 1-m_{{\rm R }z} ^2) + T_{{\rm LL}}^2
( 1-m_{{\rm L}z} ^2)}{T_{{\rm RL}}^2} ,\nonumber \\
\phi^{\pm}&=&\phi({\sf v}+{\sf b})\pm \phi({\sf v}-{\sf b}) ,
\label{hBgenP}
\end{eqnarray}
provided that we do not have one of the cases $m_{{\rm R} z}=m_{{\rm L}
z} =\pm 1$. In the latter two cases, the self-consistency equation is
an identity and the spin steady state can be any. When ${\bf
m}_{\alpha}=0$ we reproduce the results of Ref.~\cite{ParcolletHooley}
for unpolarized electrons (see Fig.~\ref{fig5}).  If we account for
environment-driven relaxation, $\Gamma_{{\sf env}}\ne 0$, the
distribution function  will be closer to that in thermal equilibrium.
When such  relaxation dominates,  the distribution function
$h_f(\omega)$ coincides with the equilibrium one.

\subsection{Average spin magnetization and the spin-spin correlation 
function}
\label{sec4c}

We compute first the average spin magnetization $\langle
S_\mu(t)\rangle=-(i/2) \epsilon^{\mu\nu\gamma}\langle
\eta_\nu(t)\eta_\gamma(t)\rangle$ (time-independent in the dc steady
state),  where $\epsilon^{\mu\nu\gamma}$ is the antisymmetric unit
tensor.  To lowest order, the values $\langle S_x\rangle$ and $\langle
S_y\rangle$ are proportional to  the small parameters
$T_{\alpha\beta}^2\rho_0^2$ and can be neglected.  For $\langle
S_z\rangle$, we get (to zeroth order)
\begin{equation}
\langle S_z\rangle=-\frac{1}{2} h_f(B) .
\end{equation}
When ${\bf m}_{{\rm R}}={\bf m}_{{\rm L}}$ (except $m_{\alpha z}=\pm 1$),
we obtain the same result as for unpolarized electrons found by
Parcollet and Hooley
\begin{equation}
\langle S_z\rangle=\frac{1}{2}\tanh\left(\frac{\sf b}{2}\right) \
\frac{\phi({\sf b})(2+\theta)}
{\phi^++\phi({\sf b})\theta}.
\end{equation}
Here $\theta=(T_{{\rm LL}}^2+T_{{\rm RR}}^2)/T_{{\rm RL}}^2$. At $T=0$,
and $V<B$ we get $\langle S_z\rangle=1/2$, i.e., the spin is in the
ground state  because electrons do not have enough energy to flip the
spin. When $V>B$ the electrons can flip the spin, thus  reducing the
spin magnetization to $\langle S_z\rangle = (1+\theta/2)/(2{\sf
x}+\theta)$, where ${\sf x}=V/B$. It drops as $1/V$ for large $V$.

As noticed by Shnirman and Makhlin \cite{Shnirman} the spin-spin
correlation function ${\cal S}_{xx}(t)=\langle S_x(t)S_x(0)\rangle$
may be expressed via the $f$-fermion Green's functions.
We rewrite the spin operators in the following fashion
\begin{eqnarray}
S_x&=&-(f^{\dagger}+f)\hat{\tau}/2= -\hat{\tau}
(f^{\dagger}+f)/2 , \nonumber \\
S_y&=&i(f^{\dagger}-f)\hat{\tau}/2= i\hat{\tau}
(f^{\dagger}-f)/2 , \label{tran}\\
S_z&=&-\eta_z\hat{\tau}/\sqrt{2}= - \hat{\tau}
\eta_z/\sqrt{2} .\nonumber
\end{eqnarray}
The operator
$\hat{\tau}=\sqrt{2}(1-2f^{\dagger}f)\eta_z=i2\sqrt{2}\eta_x\eta_y\eta_z$
commutes with the spin operators $S_\mu$ and hence with the
Hamiltonian, while $\hat{\tau}^2=\one$. (Notice that the operator
$\hat{\tau}$ maps the orthogonal subspaces  ${\cal H}_{1,\pm}
\leftrightarrow {\cal H}_{2,\pm}$ of section \ref{3} among themselves.)
As a result, 
\begin{eqnarray}
{\cal S}_{xx}(t)&=&\langle S_x(t)S_x(0)\rangle \label{tr1} \\
&=&\frac{1}{4}\langle  [f^{\dagger}(t)+f(t)][f^{\dagger}(0)+f(0)]
\rangle   \nonumber \\
&=&\frac{i}{4}[G_{ff^{\dagger}}^{-+}(t)+G_{f^{\dagger}f}^{-+}(t)+
G_{f^{\dagger}f^{\dagger}}^{-+}(t)+G_{ff}^{-+}(t)].\nonumber
\end{eqnarray}
Similarly,
\begin{eqnarray}
{\cal S}_{yy}(t)&=&-\frac{1}{4}
\langle  [f^{\dagger}(t)-f(t)][f^{\dagger}(0)-f(0)]\rangle
 \nonumber \\
&=&\frac{i}{4}[G_{ff^{\dagger}}^{-+}(t)+G_{f^{\dagger}f}^{-+}(t)-
G_{f^{\dagger}f^{\dagger}}^{-+}(t)-G_{ff}^{-+}(t)].
 \nonumber \\
{\cal S}_{zz}(t)&=& \frac{i}{2} \ G_{zz}^{-+}(t). \label{szz}
\end{eqnarray}
and, in general, ${\cal
S}_{\mu\nu}(t)=\frac{i}{2}G_{\mu\nu}^{-+}(t)$.

To zeroth order in perturbation theory we obtain
\begin{eqnarray}
\label{sxxo}
&&{\cal S}_{xx}(\omega)={\cal S}_{yy}(\omega)=
\frac{i}{4}[G_{ff^{\dagger}}^{-+}(\omega)+
G_{f^{\dagger}f}^{-+}(\omega)] \label{bare} \\
&=&\frac{\pi}{4}
\{[1-h_f(B)]\delta(\omega -B)+[1+h_f(B)]\delta(\omega +B)\},
 \nonumber
\end{eqnarray}
while ${\cal S}_{zz}(\omega)=(\pi/2) \ \delta(\omega)$.

To describe the contribution which displays a peak at the Larmor frequency
in the correlation function ${\cal S}_{xx}(\omega)$ we need
to account for the imaginary part of the self-energy.
For $G_{ff^{\dagger},f^{\dagger}f}$ we have
\begin{eqnarray}
G_{ff^{\dagger}}^{R,A}(\omega)&=& \frac{1}{\omega -B -
\Sigma _{f^\dagger f}^{R,A}} \ , \\
G_{f^\dagger f}^{R,A}(\omega)&=& \frac{1}{\omega +B -\Sigma
_{ff^\dagger}^{R,A} } \ . \nonumber
\end{eqnarray}
At the weak tunneling condition, $|T_{{\rm RL}}|^2\rho_0^2\ll
1$, when the relaxation due  to the environment is negligible, we can
also neglect  the contributions from $G_{ff}$ and $G_{f^\dagger
f^\dagger}$ to ${\cal S}_{xx}$  with respect to  those from $G_{f
f^\dagger}$ and $G_{f^\dagger f}$. We obtain
\begin{equation}
{\cal S}_{xx}(\omega)= \frac{1}{4}
\left[ \frac{(1-h_f(B))\Gamma_{\perp} }{(\omega - B)^2 + \Gamma_{\perp}
^2}+ \frac{(1+h_f(B))\Gamma_{\perp}}{(\omega + B)^2 + \Gamma_{\perp}
^2}\right],
\label{gamma}
\end{equation}
where, at $T=0$, for fully polarized electrons along the $x$-axis 
we get
\begin{eqnarray}
\!\!\!\!\!\Gamma_{\perp}(B,V)&=&{\rm Im}\Sigma_{f^\dagger f}^A(B) \nonumber \\
&=&\frac{\pi}{4}T^2_{{\rm RL}} \rho_0^2 [
|V+B|+|V-B|+\theta B]. 
\label{gammasxx}
\end{eqnarray}
When $V<B$ this gives the width of the precession resonance due to
quantum fluctuations of the current (the current operator does not
commute with the Hamiltonian, and the current fluctuates between the
leads and the spin),
\begin{equation} 
\Gamma_{\perp}^{(0)}(B)=\frac{\pi}{4}T_{{\rm RL}}^2\rho_0^2 B(2+\theta),
\end{equation}
while for $V>B$ the decoherence rate has an additional contribution due
to the voltage induced transport current 
\begin{equation}
\Gamma_{\perp}(B,V)=\Gamma_{\perp}^{(0)}(B)+
\frac{\pi}{2} T_{{\rm RL}}^2\rho_0^2 (V-B) \Theta(V-B).
\end{equation}
We estimate $\Gamma_{\perp}\approx 10^{-5}V$, in a junction with $T_{\rm RL}
=0.1T_0$ and tunneling resistance $R_T=100$ Mohm. To
determine $\Gamma_{\perp}$ in the unpolarized case we need to put 
${\bf m}_{\alpha}=0$ and multiply by 4 in Eq. (\ref{gammasxx}). 

Thus, we see that the spin-spin correlation functions ${\cal
S}_{xx}(\omega)$ and  ${\cal S}_{yy}(\omega)$ exhibit oscillations at a
renormalized Larmor frequency $\omega_L$, with a peak width depending
on $V$, $B$ and $T$.  When the relaxation due to the environment
dominates, $B\gg\Gamma_{{\sf env}}\gg  \Gamma_{\perp}$, we need to
replace  $\Gamma_{\perp}$ by $\Gamma_{{\sf env}}$ in Eq.~(\ref{gamma}).
The important point is that $\Gamma_{\perp}$ provides a lower bound for
the decoherence rate $\Gamma$, i.e., for the width of the peak at
$\omega_L$ in the spin-spin correlation function.


\section{Effect of the spin on the tunneling current}
\label{sec5}
\subsection{The average current}
\label{sec5a}

Let us now turn to the evaluation of the average current and current
power spectrum.  The current operator is given as
\begin{equation}
\hat{I}(t)=-ie\!\!\!\sum_{n,n',\sigma,\sigma'}
[c_{{\rm R}n\sigma}^{\dagger}(t)(\hat{T}_{{\rm RL}})_{\sigma\sigma'}
c^{\;}_{{\rm L}n'\sigma'}(t) -{\rm H}.{\rm c}.] .
\end{equation}

\begin{figure}
\centerline{\epsfxsize=6cm \epsfbox{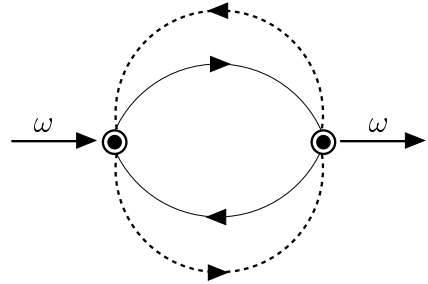}}
\caption{A diagram for the current-current correlation function
that does not contribute to the peak at the renormalized Larmor frequency
but gives a spin-dependent correction to the shot noise.}
\label{fig6}
\end{figure}

The average current (involving the complete spin-electrons Hamiltonian
${\cal H}$) is determined from
\begin{equation}
I=  \langle \hat{I}(t)\rangle=2e \ {\rm
Im}\sum_{n,n',\sigma,\sigma'}\!\!\!\langle c_{{\rm R}n
\sigma}^{\dagger}(t)(\hat{T}_{{\rm RL}})_{\sigma\sigma'}
c^{\;}_{{\rm L}n'\sigma'}(t)\rangle.
\end{equation}
In the following, we will assume that ${\cal H}_T$ is turned on
adiabatically and we are interested only in the average current up to
second order in ${\cal H}_T$. (Remember that we are only interested in
the dc steady state solution.) The following derivation is similar to
that explained in Ref.~\cite{Mahan} for the quantum dot tunneling.
Denoting
\begin{equation}
\hat{A}(t)= \!\!\sum_{n,n'
\sigma,\sigma'}\!\! \ c_{{\rm R}n
\sigma}^{\dagger}(t)(\hat{T}_{{\rm
RL}})_{\sigma\sigma'}c^{\;}_{{\rm L}n'
\sigma'}(t) ,
\end{equation}
we obtain (assuming that the leads are not superconducting)
\begin{eqnarray}
\langle \hat{I}(V) \rangle &=&-2e \ {\rm Im}\{U^{R}(-V)\}, \\
U^{R}(\omega)&=&-i\int_{-\infty}^{\infty} \!\!\! dt \ e^{i\omega
t}\theta(t) \ \langle[\hat{A}(t),\hat{A}^{\dagger}(0)]\rangle.
\nonumber
\end{eqnarray}
Assuming the dependence $\rho(\epsilon)=\rho_0$ near the Fermi energy
we obtain for fully polarized electrons, $B_{\alpha}\gg D$,
\begin{eqnarray}
I(V) &=&
I_0(V)+{\bf I}_s(V)\cdot \langle{\bf S}\rangle, \label{avcur} \\
I_0(V)&=&\pi e (1+{\bf m}_{\rm R}\cdot{\bf m}_{\rm L}) \
T_0^2\rho_0^2 \ V, \nonumber\\
{\bf I}_{s}^{(1)}(V)&=&2\pi e ({\bf m}_{{\rm
R}}+{\bf m}_{{\rm L}}) \ T_0T_{{\rm RL}}^{({\sf ex})}
\rho_0^2 \ V.\nonumber
\end{eqnarray}
For weakly polarized electrons, $B_{\alpha}=B\ll D$,  approximating
$\rho(\epsilon)\approx\rho_0(1+\epsilon/D)$ near the Fermi energy, we
get $|{\bf I}_s|\propto e T_0T_{{\rm RL}}^{(\sf ex)}\rho_0^2 (B/D) V$,
while $I_0=4\pi e T_0^2\rho_0^2 V$.

When $({\bf m}_{{\rm R}},{\bf m}_{{\rm L}})\perp {\bf B}$  the term of
order $T_0T_{{\rm RL}}^{({\sf ex})}$ vanishes because $\langle {\bf
S}\rangle$  is parallel to $ {\bf B}$. The next spin-dependent term in
the  average current, $I_{s\perp}^{(2)}$, is of order  $(T_{{\rm
RL}}^{({\sf ex})})^2$ and, in the fully polarized case along the
$x$-axis, is given by
\begin{eqnarray}
I_{s\perp}^{(2)}&=&2e(T_{{\rm RL}}^{({\sf ex})})^2{\rm Re}\{\int_{-\infty}^t
dt'e^{-iV(t-t')} \nonumber \\
&&[G_{{\rm R}}(t'-t)G_{{\rm L}}(t-t') \ {\cal S}_{xx}(t-t')-\\
&&G_{{\rm R}}^*(t'-t)G_{{\rm L}}^*(t-t') \ {\cal S}_{xx}(t'-t)]\}. \nonumber
\end{eqnarray}
Here $G_\alpha(t)$ is the causal Green's function. Using Fourier
transforms and Eq.~(\ref{sxxo}) the result, at $T=0$, is
\begin{equation}
I_{s\perp}^{(2)}=\frac{\pi}{2} e  (T_{{\rm RL}}^{({\sf ex})})^2
\rho_0^2 \ [V+h_f(B) B] \ \Theta(V-B).
\end{equation}
Hence, when $V<B$ the average current is not affected by the spin
because electrons have not enough energy to flip it. The spin remains
in the ground state and thus cannot affect tunneling electrons and be
probed. In the case that there is no reflection, i.e., $\theta=0$, the
spin-dependent contribution is related to  the spin-flip rate
$\Gamma_{\parallel}$,
\begin{equation}
I_{s\perp}^{(2)}=e \Gamma_{\parallel} , 
\end{equation}
because the spin-dependent contribution to the current is determined by
the matrix element of the operator $S_x$ between states with opposite
spin projections. Hence, each tunneling electron needs to flip the
localized spin in order to go through. In the large $V$ limit we obtain
$\Gamma_{\parallel}\rightarrow \Gamma_{\perp}$ in agreement with the
quasiclassical results  \cite{KorotkovAverin,BulaevskiiOrtiz}.

At $T=0$, the contribution $I_{s\perp}^{(2)}$ leads to a discontinuous 
derivative of $I(V)$  at $V=B$ due to the opening of a new channel
corresponding to the tunneling  of electrons via the spin. The change
is more pronounced in the second order derivative of $I(V)$  with
respect to $V$. Accounting for broadening  of the spin energy,
$\Gamma_{\perp}$, we derive
\begin{equation}
\frac{d^2I}{dV^2}=2e(T_{{\rm RL}}^{({\sf ex})})^2\rho_0^2
\sum_{\mu=x,y}{\cal S}_{\mu\mu}(V)(m_{{\rm R}\mu}+m_{{\rm L}\mu})^2.
\label{peak}
\end{equation}
We see that this second order derivative of the average tunneling
current  shows a peak (see Eq.~(\ref{sxxo})) at $V=B$.  Hence,
measurements of the $I$-$V$ tunneling characteristics provide
information on the spectral density of the spin-spin correlation
function.  Similar behavior for the average current with respect to
bias voltage was  used previously for inelastic tunneling spectroscopy
of phonons \cite{Stipe}.  It is based on an equation similar to
Eq.~(\ref{peak}) relating $d^2I/dV^2$  to the phonon density of
states.

Note, that in the case of unpolarized electrons $I_s$ is small, but
the second  derivative of $I(V)$ also shows a similar peak \cite{Ab}. 
However, now
its amplitude has an additional small factor  $T_0^2\rho_0^2$.

\subsection{The shot noise power}
\label{sec5b}

Next we calculate the current-current correlation function.  The
particular fluctuation operator one needs to evaluate depends upon the
nature of the measurement itself \cite{lesovik}, see also \cite{Kogan}.
In the general case the result of measurements depends on the
correlation functions describing emission, (+), and absorption, (-), by
the tunneling contact 
\begin{equation}
{\cal P}_{\pm}(\omega) =\int _{-\infty}
^{\infty} \!\!\!dt \ e^{\pm i \omega t} \ {\cal P}(t), \ \ \ 
{\cal P}(t)= \langle\hat{ I}(0)\hat{ I}(t)\rangle.
\label{ii}
\end{equation}
We obtain
\begin{eqnarray}
&&{\cal P}_{\pm}(\omega)={\cal P}_{s\pm}(\omega)+{\cal
P}_{p\pm}(\omega), \label{ps} \\
&&{\cal P}_{s}(t)=
e^2 \!\!\!\!\!\sum_{n,n',
m,m',\sigma,\sigma',\gamma,\gamma'}\!\!\!\!\!\!\!\!\!\!\!\![
\langle\langle c_{{\rm R}n\sigma}^{\dagger}(0)c^{\;}_{{\rm
R}n'\gamma}(t)\rangle_e
(\hat{T}_{\rm RL})_{\sigma\sigma'}(0) \nonumber \\
&&\times\langle c^{\;}_{{\rm L}m\sigma'}(0)c_{{\rm L}m'
\gamma'}^{\dagger}(t)\rangle_e(\hat{T}_{\rm
RL})_{\gamma'\gamma}(t)\rangle_s+({\rm R }\leftrightarrow {\rm L})],
\nonumber
\end{eqnarray}
where the term ${\cal P}_s(t)$ describes the shot noise to second
order  in the matrix elements $T_0$ and $T_{{\rm RL}}$, while  ${\cal
P}_p(t)$ represents the contribution which displays a peak at the
renormalized Larmor frequency of order $(T_0T_{{\rm RL}})^2$  (see next
subsection). We calculate ${\cal P}_s(t)$ in the  framework of standard
time-dependent perturbation theory (as we did to  calculate the average
current). In the equation above, $\langle ...\rangle_e$ means average
over electron degrees of freedom, after the perturbation due to the
spin-electron interaction is accounted for, while $\langle
...\rangle_s$ means average over the localized spin degrees of freedom.
Averages over spin operators are performed using the known spin
correlation functions determined in previous sections. 

Contributions proportional to $T_0^2$ and $T_0T_{{\rm RL}}$ give the
standard expression \cite{Scalapino} modified by the presence of the
localized spin, in the same way as the average current
Eq.~(\ref{avcur}). For fully  polarized electrons along the $x$-axis we
obtain, at $T=0$,
\begin{equation}
{\cal P}_{s+}^{(0)}(\omega)=2\pi e^2 (T_0^2+2T_0T_{{\rm RL}})
\rho_0^2 \ |V-\omega|. \label{shot}
\end{equation}
For unpolarized electrons we need to multiply this result by a factor
4. We obtain for the symmetrized shot noise power ($\omega<V$)
\begin{equation}
[{\cal P}_{s+}^{(0)}+{\cal P}_{s-}^{(0)}]/2=e(I_0+I_s^{(1)}).
\end{equation}
The terms proportional to $T_{{\rm RL}}^2$ describe the effect of the
spin dynamics on the shot noise, ${\cal P}_s^{(1)}$. A diagram
contributing to ${\cal P}_s^{(1)}$ is shown in Fig. \ref{fig6}.
For electrons fully polarized along the $x$-direction we derive
for the emission part at $T=0$ and $\Gamma_{\perp}=0$
\begin{eqnarray} 
{\cal P}_{s+}^{(1)}(\omega)&=& \frac{\pi}{4} e^2(T_{{\rm RL}}^{({\rm
ex})})^2\rho_0^2 \ \Theta(V-B) [(1-h_f(B)) \label{spinn} \\
\!\!\!\!\!\!\times&&\!\!\!\!\!\!\!\!\!\!\!\!F(-\omega+V-B)
+(1+h_f(B))F(-\omega+V+B)], \nonumber
\end{eqnarray}
where 
\begin{equation}
F(x)=x \ \Theta(x).
\end{equation}
Emission is possible at frequencies $\omega<V-B$, when the spin is in
the  ground state, and frequencies $\omega<V+B$, when the spin is in an
excited state. The second order derivative of ${\cal
P}_{s+}^{(1)}(\omega)$ with respect to $\omega$, when $V>B$,  has peaks
with amplitudes $\pi e^2(T_{{\rm RL}}^{({\rm ex})})^2\rho_0^2(1\pm
h_f(B))/2$ and widths $\Gamma_{\perp}$  at frequencies $\omega=V\pm B$.
This provides a way to detect the presence of the localized spin.

\begin{figure}
\centerline{\epsfxsize=7.5cm \epsfbox{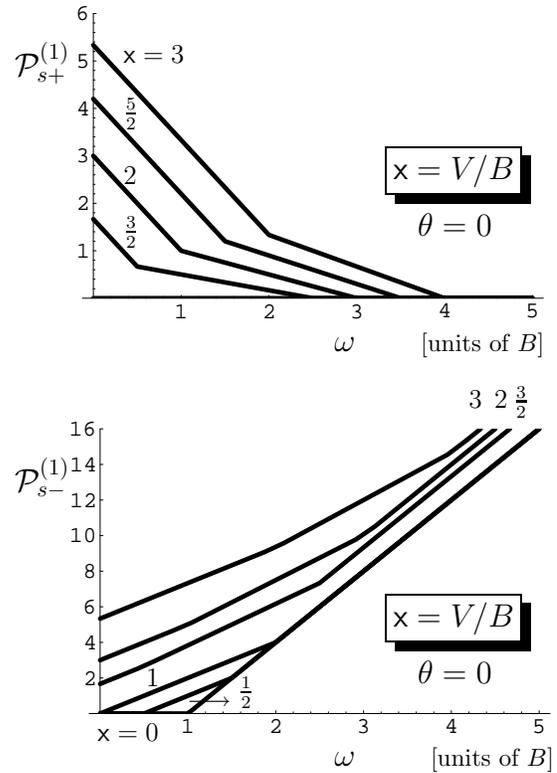}}
\caption{Emission and absorption contributions to the shot noise power
to order $T_{{\rm RL}}^2$, at $T=0$, for similar electron
polarizations in the leads ${\bf m}\perp {\bf B}$. ${\cal
P}_{s\pm}^{(1)}$ is in units  $(\pi Be^2(T_{{\rm RL}}^2\rho_0^2/4)$. 
For ${\sf x}<1$ the emission contribution
vanishes. Notice the different regimes, characterized by a change in
the slope, that appear because of the presence of the localized spin.
} 
\label{fig7}
\end{figure}

The absorption part at $T=0$ and $\Gamma_{\perp}=0$ is given as 
\begin{eqnarray}
\!\!\!\!\!\!\!{\cal P}_{s-}^{(1)}(\omega)&=&\frac{\pi}{4} e^2 
(T_{{\rm RL}}^{({\rm ex})})^2 \rho_0^2 \ \{ \nonumber \\
\Theta(V-B)&&\!\!\!\!\!\!\!\!\!\!\![(1-h_f(B))(\omega+V-B+F(\omega-V-B))
 \nonumber \\
&&\!\!\!\!\!\!\!\!\!\!+(1+h_f(B))(\omega+V+B+ F(\omega-V+B))]  \nonumber \\
+2\Theta(B-V) &&\!\!\!\!\!\!\!\!\! [ F(\omega+V-B)+F(\omega-V-B) ] \ \}
,
\label{spinnn}
\end{eqnarray}
with the result 
\begin{equation}
[{\cal P}_{s+}^{(1)}+{\cal P}_{s-}^{(1)}]/2=e I^{(2)}_{s \perp}
\end{equation}
for $\omega < |V-B|$. Figure \ref{fig7} displays the emission and
absorption contributions. 

For unpolarized electrons we derive 
\begin{equation}
{\cal P}_{s+}^{({\rm un},1)}(\omega)=4{\cal P}_{s+}^{(1)}(\omega)+
2\pi e^2(T_{{\rm RL}}^{({\rm ex})})^2\rho_0^2 \ (V-B),
\label{spinun}
\end{equation}
where ${\cal P}_{s+}^{(1)}(\omega)$ is given by Eq.~(\ref{spinn}) with
the  same $h_f(B)$ as for fully polarized electrons. This noise power
also exhibits anomalies at frequencies $\omega=V\pm B$. We see
that ${\cal P}_s^{(1)}(\omega)$ has no peak at the Larmor frequency
neither for polarized nor for unpolarized electrons contrary to the
results of Ref.~\cite{BalatskyManassen}.

\begin{figure}
\centerline{\epsfxsize=8.5cm \epsfbox{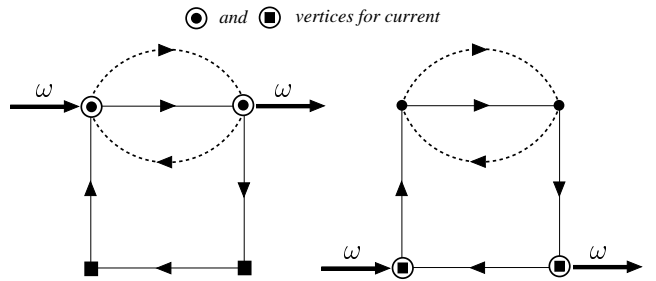}}
\caption{Two of the six diagrams for the current-current correlation
function that do not contribute to the peak at the renormalized Larmor
frequency to order $T_0^2T_{{\rm RL}}^2$. The external vertices can be
any two of the four vertices in the diagram.}
\label{fig8}
\end{figure}

The fourth order terms, $T_0^2T_{{\rm RL}}^2$, in ${\cal P}(\omega)$
correspond to 10 skeleton diagrams some of which are shown in Figs.
\ref{fig8} and \ref{fig9}. In these 4-vertex diagrams the  external
frequency $(\omega)$ (via the current vertex) may enter in any of the
two vertices. Diagrams in Fig. \ref{fig8} (all six), as well as the
diagram shown in Fig.~\ref{fig6}, contain integration over the energy
$\epsilon$ entering in all the electron and spin lines of the diagram.
Hence, these contributions change at least on the frequency scale of
order $V,B$ as the term ${\cal P}_s^{(1)}(\omega)$. They are small in
comparison to ${\cal P}_s^{(1)}(\omega)$, and may be neglected.

\subsection{The peak at the renormalized Larmor frequency}
\label{sec5c}

\begin{figure}
  \centerline{\epsfxsize=6cm \epsfbox{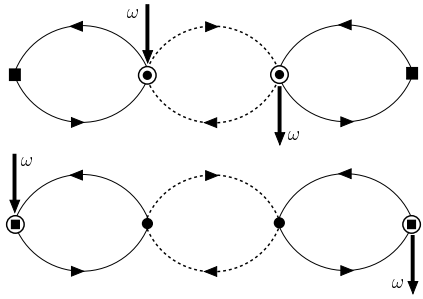}}
  \caption{Two of the four diagrams contributing to the peak at the renormalized Larmor frequency in the current-current
correlation function to order $T_0^2T_{{\rm RL}}^2$. There are two other diagrams with different current
vertices. }
  \label{fig9}
\end{figure}

To derive this peak we consider the 4 diagrams of the type shown  in
Fig. \ref{fig9}, which give contributions of order $T_0^2T_{{\rm
RL}}^2$. In these diagrams integration over energy $\epsilon$ in the
spin lines is independent of the integrations in the electron lines and
they provide a contribution to the current-current correlation function
${\cal P}_p(\omega)$  which has the same peaks at $\omega=\pm B$ as the
spin-spin correlation function. To calculate this contribution we use
the Keldysh technique as done by Shnirman {\it et al.} \cite{Shnirman}.
We will assume that to study the peak at the renormalized Larmor
frequency experimentally one can measure the instantaneous values of
the current over a long period of time and then obtain the symmetrized
current power spectrum defined as
\begin{equation}
{\cal P}(\omega) =\frac{1}{2} [{\cal P}_+(\omega)+{\cal P}_-(\omega)].
\label{si}
\end{equation}
To calculate the first term in Eq.~(\ref{ii}), we put $t=t_-$ and
$0=0_+$. For electrons fully polarized along the $x$-axis the first term
is given as
\begin{eqnarray}
&&\langle I(t)I(0)\rangle=\langle-T_0^2{\hat B}(t){\hat B}(0)
-T_0T_{\rm RL}[{\hat A}(t)S_x(t){\hat B}(0) \nonumber \\
&&+{\hat B}(t){\hat A}(0)S_x(0)]
-T_{\rm RL}^2{\hat A}(t)S_x(t){\hat B}(0)S_x(0)\rangle, 
\label{3t} \\
&&{\hat B}(t)=\sum_n c_{{\rm R}nx}^+(t)c_{{\rm L}nx}(t), \nonumber \\
&&{\hat A}(t)=\sum_n c_{{\rm R}nx}^+(t) \sigma _x c_{{\rm L}nx}(t).
 \nonumber
\end{eqnarray}
Now, in the framework of the Keldysh technique,
we expand $B(t)B(0)$ up to $(T_{\rm RL})^2$ order in the first term,
up to $T_0T_{\rm RL}$
in the second and third terms and up to $T_0^2$ in the last term.
Let us consider the first term which 
corresponds to the second diagram in Fig.~\ref{fig9}. This diagram leads to 
the contribution 
$$- (T_0T_{{\rm RL}})^2\sum_{\mu,\nu=\pm}\Lambda_{\rm RL}^{-\mu}
(-\omega)\mu\nu {\cal S}^{\mu\nu}(\omega)\Lambda^{\nu+}(\omega).$$
Here the product $\mu\nu$ is the sign factor
$(+1)$ if $\mu=\nu$ and $(-1)$ otherwise,
while
\begin{eqnarray}
&&\Lambda_{\rm RL}^{\mu\nu}(\omega)=\int\frac{d\epsilon}{2\pi}
{\rm Tr} [{\bf G}_{\rm R}^{\mu\nu}(\epsilon)\sigma_x{\bf G}_{\rm L}^{\nu\mu} (\epsilon-\omega)] \ ,\\
&&\Lambda^{\mu\nu}(\omega)=
[\Lambda_{\rm RL}^{\mu\nu}(\omega)-\Lambda_{\rm LR}^{\mu\nu}(\omega)], 
\end{eqnarray}
where the trace is taken over spin variables.
Further, we denote by ${\cal S}^{\mu\nu}(\omega)$ the combination of Majorana fermion
Green's functions which are represented by broken lines in the diagrams in
Fig.~\ref{fig9},
\begin{eqnarray}
&&{\cal S}^{\mu\nu}(\omega)= \int \frac{d\epsilon}{2\pi}
G_{yy}^{\mu\nu}(\epsilon+\omega)G_{zz}^{\nu\mu}(\epsilon)=
\label{rep}\\
&&\int \frac{d\epsilon}{2\pi}
G_{zz}^{\mu\nu}(\epsilon+\omega)G_{yy}^{\nu\mu}(\epsilon)=
\int \frac{d\epsilon}{2\pi}
G_{zz}^{\mu\nu}(\epsilon)G_{yy}^{\nu\mu}(\epsilon-\omega), \nonumber
\end{eqnarray}
with the property
\begin{equation}
{\cal S}^{\mu\nu}(\omega)={\cal S}^{\nu\mu}(-\omega).
\end{equation}
Adding the term with interchanged R and L and symmetrizing with respect to
time we obtain for the first term
\begin{eqnarray}
&&{\cal P}_p^{(1)}(\omega)= -\frac{1}{2}(T_0T_{{\rm RL}})^2
\sum_{\mu,\nu}\mu\nu
[\Lambda^{-\mu}(\omega)
{\cal S}^{\mu\nu}(\omega)\Lambda^{\nu+}(\omega) \nonumber \\
&&+  \Lambda^{-\mu}(-\omega)
{\cal S}^{\mu\nu}(-\omega)\Lambda^{\nu+}(-\omega)],
\end{eqnarray}
Similarly we calculate the other three terms
\begin{eqnarray}
&&{\cal P}_p^{(2)}(\omega)= -\frac{1}{2}(T_0T_{{\rm RL}})^2
\sum_{\mu,\nu}\mu\nu
[\Lambda^{-\mu}(0){\cal S}^{-\nu}(\omega)\Lambda^{\nu+}(\omega)
 \nonumber\\
&&+\Lambda^{-\mu}(0){\cal S}^{-\nu}(-\omega)\Lambda^{\nu+}(-\omega)], \\
&&{\cal P}_p^{(3)}(\omega)=-\frac{1}{2}(T_0T_{{\rm RL}})^2
\sum_{\mu,\nu}\mu\nu[\Lambda^{-\mu}(\omega)
{\cal S}^{\mu-}(\omega)\Lambda^{\nu+}(0) \nonumber\\
&&+\Lambda^{-\mu}(-\omega){\cal S}^{\mu-}(-\omega)\Lambda^{\nu+}(0)], \\
&&{\cal P}_p^{(4)}(\omega)=-\frac{1}{2}(T_0T_{{\rm RL}})^2
\sum_{\mu,\nu}\mu\nu
\Lambda^{-\mu}(0)[{\cal S}^{-+}(\omega)+ \nonumber \\
&&S^{-+}(-\omega)]\Lambda^{\nu+}(0).
\end{eqnarray}
Let us calculate ${\cal S}^{\mu\nu}(\omega)$ using the expressions for $G^{-+}$ and
$G^{+-}$ in terms of the zeroth-order functions $G^R$, $G^A$ and $G^K$ 
\begin{eqnarray}
&&{\cal S}^{-+}(\omega)=\frac{1}{4}\int\frac{d\epsilon}{2\pi}
\{G_{yy}^K(\epsilon+\omega)G_{zz}^K(\epsilon)+ \\
&&G_{yy}^A(\epsilon+\omega)G_{zz}^R(\epsilon)+
G_{yy}^R(\epsilon+\omega)G_{zz}^A(\epsilon)+ \nonumber \\
&&G_{zz}^K(\epsilon+\omega)[G_{yy}^R(\epsilon)-G_{yy}^A(\epsilon)]+
\nonumber \\
&&G_{yy}^K(\epsilon+\omega)[G_{zz}^A(\epsilon)-G_{zz}^R(\epsilon)]\}.
\nonumber
\end{eqnarray}
Since $h_z(0)=0$, we have $G_{zz}^K(\omega)=0$. Therefore, the contribution
to $S^{-+}(\omega)$ from the terms containing $G_{zz}^K$ is negligible, but
the last term does contribute. We get
\begin{eqnarray}
&&{\cal S}^{-+}(\omega)={\cal S}^{+-}(-\omega) \\
&&=\frac{\pi}{4}\{[1-h_f(\omega)]\delta(\omega -B)
+[1+h_f (- \omega)] \delta (\omega +B) \}, \nonumber\\
&&{\cal S}^{++}(\omega)=\frac{1}{2}ih_f(\omega)
{\rm P}\frac{\omega}{\omega^2-B^2}+\frac{\pi}{4}\delta^+(\omega), \\
&&{\cal S}^{--}(\omega)=-\frac{1}{2}ih_f(\omega){\rm P}
\frac{\omega}{\omega^2-B^2}+\frac{\pi}{4}\delta^+(\omega),
\end{eqnarray}
where $\delta^+(\omega)=\delta(\omega-B)+\delta(\omega+B)$.
Next, we take into account the broadening of the resonances at 
$\omega=\pm B$ by replacing
\begin{equation}
\pi\delta(\omega\pm B)\Rightarrow \frac{\Gamma_{\perp}}
{(\omega\pm B)^2+\Gamma_{\perp}^2}.
\end{equation}
Such procedure follows from the fact that accounting for higher order terms
of perturbation theory in $T_{{\rm RL}}$ we need to replace the
product $G_{yy}^{\mu\nu}(t)G_{zz}^{\nu\mu}(-t)$
in Eq.~(\ref{rep}) by $-\langle \eta_y(t_{\mu})\eta_y(0_{\nu})\eta_z(0_{\nu})
\eta_z(t_{\mu})\rangle$. This function has the same frequency dependence
as $G_{ff^{\dagger}}(t)$ because from one hand
\begin{equation}
\langle S_x(t)S_x(0)\rangle=-\langle \eta_y(t)\eta_z(t)\eta_y(0)\eta_z(0)
\rangle,
\end{equation}
while from the other this spin-spin correlation function is determined by
Eq.~(\ref{tr1}).
At $T=0$ we obtain, neglecting terms of order $\omega/D$, $V/D$,
\begin{eqnarray}
&&\Lambda_{\rm RL}^{-+}(\omega,V)=\Lambda_{\rm LR}^{-+}(\omega,-V)=
2 \pi(\omega+V)\Theta(\omega+V) \rho_0 ^2, \nonumber \\
&&\Lambda_{\rm RL}^{+-}(\omega,V)=\Lambda_{\rm LR}^{-+}(\omega,-V)=
\nonumber \\
&&-2\pi(\omega+V)\Theta(-\omega-V) \rho_0 ^2, \nonumber \\
&&\Lambda_{\rm RL}^{++}(\omega,V)=\Lambda_{\rm LR}^{++}(\omega,-V)=
\Lambda_{\rm RL}^{--}(\omega,V)= \nonumber \\
&&\Lambda_{\rm LR}^{--}(\omega,-V)=\pi|\omega+V| \rho_0 ^2,
\end{eqnarray}
Summing up all terms for ${\cal P}_p(\omega)$ we obtain zero for the
amplitude of the peak at $V<B$. This same result was obtained by
Shnirman {\it et al.} \cite{Shnir}. It means that quantum fluctuations
of the spin in the ground state which are present in the spin-spin
correlation function cannot be probed by tunneling electrons.

When $V>B$ we obtain, taking into account the broadening of the
resonance,
\begin{eqnarray}
&&{\cal P}_p(\omega,V,B)= P(V,B)\frac{\Gamma_{\perp}}
{\Gamma_{\perp}^2+(\omega-B)^2}, \label{gr} \\
&&P(V,B)=4(\pi T_0T_{\rm RL}\rho_0^2)^2[V^2+VBh_f(B)]. \nonumber
\end{eqnarray}
More generally, the coefficient 4 is substituted by
$[{\bf m}_{R \perp}  +{\bf m}_{L \perp}] ^2 $ for fully polarized electrons
in the leads, while for weakly polarized electrons we get the factor
$|{\bf m}_{{\rm R}\perp}s_{{\rm R}}+{\bf m}_{{\rm L}\perp}s_{{\rm L}}|^2$.
Accounting also for spin-orbit interaction we obtain an additional term in 
$P(V,B)$ which 
is proportional to the small factor
$[(2B_{\alpha}/D)^2+\bm{\beta}^4(l_x^2+l_y^2)]$.

We see that ${\cal P}_p(\omega)$ exhibits a peak at the renormalized
Larmor frequency with a width $\Gamma_{\perp}$ only when the electrons
are polarized in a direction which is not parallel to ${\bf B}_0$.
This peak is caused by precession of the spin excited by tunneling
electrons. Tunneling of electrons excites the spin, resulting in the 
modulation
of the
tunneling current, but they also cause decoherence of the spin precession
with a rate $\Gamma_{\perp}$.

	    Therefore, the spin precession cannot be seen in the
current-current correlation function without spin-orbit interaction
when the leads-electron polarization is parallel to ${\bf B}_0$, or
the electrons are unpolarized, despite the fact that the spin-spin 
correlation functions for the $S_x$ and $S_y$
spin components exhibit oscillations at $\omega_L$.  
The reason for this is that to observe oscillations in
${\cal P}(\omega)$ one needs to probe with the current a component of
the spin perpendicular to ${\bf B}$. For that the electrons need
to have a component of polarization ${\bf m}_{{\rm R}}$ or ${\bf
m}_{{\rm L}}$ that couples to $S_x$ or $S_y$. Spin-orbit interaction
leads to a signal when the electrons are unpolarized or ${\bf
B}_0\parallel {\bf m}_{\alpha 0}$ if ${\bf B}_0$ is not parallel to
${\bf l}$. However, the signal in that case is much weaker than in the
case of fully polarized electrons.

\subsection{The signal-to-noise ratio $R$}
\label{sec5d}

For the symmetrized signal the signal-to-noise ratio reaches its
maximum  at low temperatures when $\Gamma_{{\sf
env}}\ll\Gamma_{\perp}$, while ${\bf m}_R\parallel {\bf m}_L$, and both
are perpendicular to ${\bf B}$. Indeed, electrons with components of
the spin polarization parallel to ${\bf B}$ do not contribute to the
signal but they do enhance $\Gamma_{\perp}$. When $m_{{\rm R}\perp}= m_{{\rm
L}\perp}=1$ the relaxation rate  $\Gamma_{\perp}(V,B)$ (neglecting
relaxation due to the environment) is  determined by 
Eq.~(\ref{gammasxx}).  The signal-to-noise ratio at the peak position,
for the case of perpendicular spin polarization at $T=0$ and $V>B$, is
given by the expression
\begin{equation}
R=\frac{{\cal P}_{p}(B)}{{\cal P}_{s}^{(0)}(B)}=
\frac{P(V,B)}{2\pi VT_0^2\rho_0^2\Gamma_{\perp}(V,B)} 
=4\frac{V+h_f(B)B}{V+\theta B/2} . \label{sis}
\end{equation}
Here ${\cal P}_{p}(B)$ is the height of the peak at the renormalized
Larmor frequency in the symmetrized current-current correlation
function, while ${\cal P}_{s}^{(0)}$ is the current power spectrum for
the shot noise neglecting the smaller contribution ${\cal
P}_{s}^{(1)}$.

$R$ reaches its maximum value $R_{\sf max}=4$ for $V\gg B$. In the case
of symmetrical leads, $T_{{\rm RR}}= T_{{\rm LL}}=T_{{\rm RL}}$ we get
$\theta=2$, and the function $R(V/B)$ for $\theta=0,2$ is shown in
Fig.~\ref{fig11}.

When the spin relaxation due to the environment becomes dominant,
$\Gamma_{{\sf env}}\gg \Gamma_\perp $, the signal-to-noise ratio for
the case of fully polarized electrons is smaller by the factor
$\Gamma_{\perp}/\Gamma_{{\sf env}}\ll 1$. If the perpendicular
components of ${\bf m}_{\alpha}$ were absent, we see that $R$ becomes
as small as $\bm{\beta}^4$, while for weak polarization in the
direction perpendicular to ${\bf B}$ it is as small as
$(B_{\alpha}/D)^2$.

\begin{figure}
\centerline{\epsfxsize=8.8cm \epsfbox{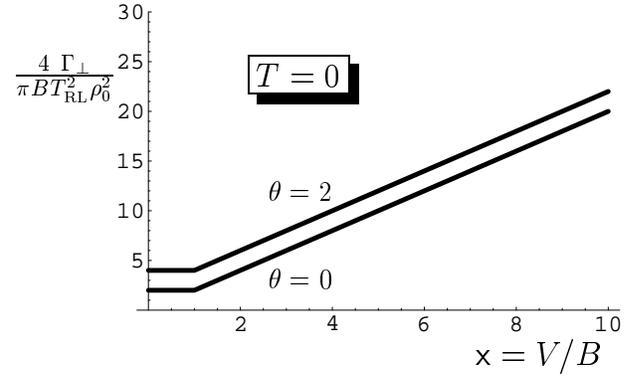}}
\caption{Width of the peak as a function of voltage for the case where
electrons are fully polarized in the direction perpendicular to the
effective magnetic field acting on the spin.}
\label{fig10}
\end{figure}
\begin{figure}
\centerline{\epsfxsize=8cm \epsfbox{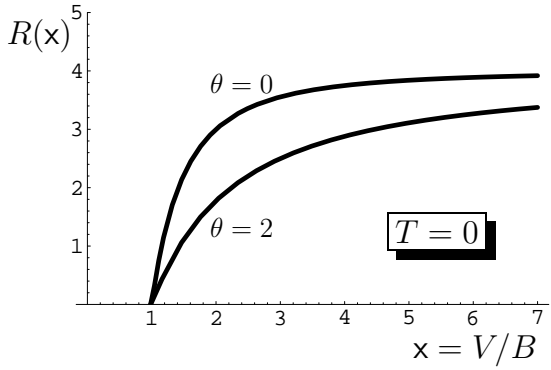}}
\caption{Signal-to-noise ratio $R({\sf x})$ for fully polarized
electrons.}
\label{fig11}
\end{figure}

\begin{figure}
\centerline{\epsfxsize=8cm \epsfbox{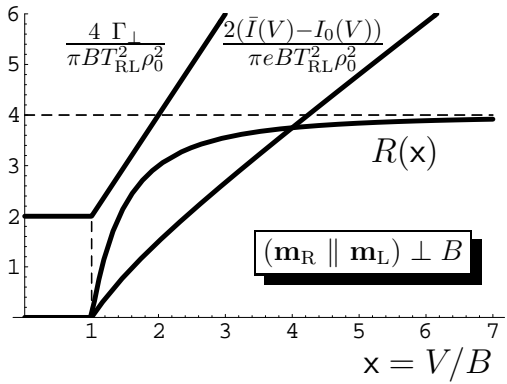}}
\caption{Comparison between the dependencies of $\Gamma_{\perp}$,
$\bar{I}-I_0$ and $R$ vs. ${\sf x}=V/B$ at $T=0$ and $\theta=0$,
explaining why $R$ reaches approaches a maximum at large $V$. }
\label{fig12}
\end{figure}

\section{Discussion}
\label{sec6}

We start summarizing our main results for the model of a localized spin
interacting with the tunneling electrons only via the exchange
interaction, i.e., when the tunneling matrix element contains the term
$T_{\rm RL}^{(\sf ex)} \vec{\bm{\sigma}}\cdot {\bf S}$ in addition to
the spin-independent term $T_0$.

{\bf 1}. The Larmor frequency is renormalized from the one determined
by the external magnetic field $B_0$ to $B=\sqrt{B_0^2+B_T^2}$ (see Eq.
(\ref{bren})), when tunneling electrons are polarized. An estimate for
$B_T$  is given after Eq.~(\ref{bren}).

{\bf 2}. In the steady state the $I$-$V$ characteristics, namely
$d^2I(V)/dV^2$, exhibits a peak at $V=B$ when the electrons are
polarized in a direction different from ${\bf B}$. This peak is due to
the opening of a new channel for electron tunneling via the localized
spin (spin-flip tunneling). The width of this peak is determined by the
spin decoherence rate $\Gamma_{\perp}$, see  Eq.(\ref{gammasxx}).

{\bf 3}. When the electrons are polarized and $V>B$, on the background 
of frequency-independent shot noise caused by the discrete nature of
the electrons, there is a peak in the current noise power $[{\cal
P}_+(\omega)+ {\cal P}_-(\omega)]/2$ at the renormalized Larmor
frequency, caused by the precession of the spin excited by the
tunneling electrons. This peak is absent when $V<B$ because the spin 
is in the ground state since electrons do not have enough energy to
flip it.

{\bf 4}. The signal-to-noise ratio for this peak,  $R={\cal
P}_p(\omega_L)/{\cal P}_s^{(0)}$, strongly depends upon the degree of
spin polarization of the leads-electrons and the orientation of this
polarization with respect to the external magnetic field ${\bf B}_0$
acting on the localized spin. The maximum value of $R$ is reached when 
the electrons are polarized in a direction perpendicular to the
effective magnetic field acting on the spin because in this case they
probe in an optimal way the precessing spin components. When the
relaxation of the localized spin with the environment is weak, $R$
reaches the values of order unity {\bf only} if the leads are almost
fully polarized in a direction close to perpendicular to ${\bf B}$.

{\bf 5}. In the case of full polarization and ${\bf m}_{\rm R0}={\bf
m}_{\rm L0}\perp {\bf B}$ at low temperatures $T\ll B_{\sf eff},V$, the
dependence of $R$ on ${\sf x}=V/B$ is given by Eq.~(\ref{sis}), where
$\theta$ comes from the leads-electron reflection via the localized
spin (see Eqs. (\ref{hBgenP})). The maximum value, $R_{{\sf max}}=4$,
is reached when $V\gg B$, see Figs.~\ref{fig11} and \ref{fig12}. The
signal-to-noise ratio is reduced by the presence of the electron
reflection couplings ($\theta \ne 0$) as expected, since this process
does not contribute to the current but does affect the width of the
peak.

{\bf 6}. When the spin relaxation due to the environment is negligible,
the peak width in the noise power spectrum (at the renormalized Larmor
frequency) depends weakly on the orientation of the polarization. It
increases linearly with $V$, when $V\gg B\gg T$ and the tunneling
electrons cause spin flips, see Fig. \ref{fig10}. At high temperatures
$T\gg V,B$, the width of the peak is determined by $T$
($\Gamma_{\perp}\propto T$).

{\bf 7}. In points 2 through 4 we assumed that the spin relaxation with
the environment, $\Gamma_{{\sf env}}$, was negligible in comparison to
the relaxation due to the leads-electrons, $\Gamma_{\perp}$. This
assumption is crucial to obtain a maximum $R$ of order unity because in
this case $\Gamma_{\perp}\propto |T_{{\rm RL}}^{(\sf ex)}|^2\rho_0^2$,
and ${\cal P}_p(B)\propto (T_{{\rm RL}}^{(\sf ex)})^2\rho_0^2
/\Gamma_{\perp}$, so that $R$  does not depend on the small parameter
$(T_{{\rm RL}}^{(\sf ex)})^2\rho_0^2$, see Eq.~(\ref{sis}). If the spin
relaxation rate due to the environment dominates we get $R\propto
\Gamma_{\perp}/\Gamma_{{\sf env}}\ll 1$.

{\bf 8}. For weak polarization of the electrons, $R$ becomes
proportional to the degree of polarization, $R\sim (B_{\alpha}/D)^2$,
where $B_{\alpha}$ is the magnetic or exchange field acting on the 
leads-electrons and $D$ represents their bandwidth. This is because
only polarized tunneling electrons contribute to the signal at the
Larmor frequency, while all tunneling electrons contribute to the
broadening of this signal. We estimate $B_{\alpha}/D\sim 10^{-5}$ for
$B_{\alpha}=100$ G and $D=0.1$ eV and this estimate leads to $R\sim
(B_{\alpha}/D)^2<10^{-8}$.

{\bf 9}. Spin-orbit (i.e., relativistic $\bm{\beta}^2$) corrections of
any type \cite{Balatsky,LevitovRashba} can only provide a very small
signal-to-noise ratio, $R\sim \bm{\beta}^4\sim 10^{-8}$, in the
situation where the exchange coupling leads to a vanishing signal
(i.e., the electrons are unpolarized or they are polarized parallel to
${\bf B}_0$). The signal in this case is determined by the relativistic
coupling between the spin and the tunneling electrons, while the
broadening is determined by the dominant non-relativistic exchange
coupling. Hence, for a tunneling process via a single localized spin in
the steady state, the signal-to-noise ratio becomes of order unity {\bf
only} when \\ 
a) the spin is decoupled from the environment, and \\ 
b) the electrons are almost fully polarized in a direction
perpendicular to the effective field acting on the localized spin.

{\bf 10}. We derived the correction to the standard shot noise power
caused by the presence of the localized spin when the leads-electrons
are unpolarized. This correction is small when $T_{{\rm RL}}\ll T_0$,
but it changes behavior at $V=B$, see Eqs.~(\ref{spinn}), and
(\ref{spinun}). On the other hand, the standard shot noise power
increases linearly with V. This new result can be observed, in
principle, experimentally.

We are now in position to compare our results to the experimental data
obtained by Manassen et al. \cite{Manassen} and Durkan and Welland
\cite{Durkan}, and decide whether the model of tunneling via a single
localized spin can explain these data. In these experiments the
polarization of the electrodes was weak, $R$ was of order unity or
larger almost independently of the orientation of the applied magnetic
field, and the position of the peak in the noise power spectrum ${\cal
P}(\omega)$ was found at the Larmor frequency corresponding to the
applied magnetic field. Our results show that a combination of weak
polarization and a signal-to-noise ratio of order unity are
incompatible for the dc steady state described by our model, which
accounts for both exchange and spin-orbit coupling of the tunneling
electrons and the localized spin.

Next, at high voltages we reproduced the quasi-classical results
\cite{KorotkovAverin,BulaevskiiOrtiz} which are valid in the limit
$V\gg B$.

From the point of view of the theory of quantum measurement we see that
after a long time, any measurement usually leads to a steady state
where information on the initial state of the quantum system is
probably lost. Hence, an important question is what is the transient
time for a given measurement.

We note that $d^2I/dV^2$ and the contribution ${\cal P}_p(\omega)$
in the current-current correlation function are proportional to the
spin-spin correlation function $S_{xx}(V)$ and $S_{xx}(\omega)$,
respectively (if the electrons are spin polarized along the $x$-axis).
Suppose that one wants to probe a single spin, say A, which is part of
an ensemble of other spins interacting with each other. Measurement of
the $I$-$V$ characteristics, namely $d^2I/dV^2$, and current noise
power, ${\cal P}(\omega)$, at low voltages will carry information on
the dynamics of that particular (coupled) spin A, whenever tunneling
occurs via that single spin. Effectively, because of the coupling, that
measurement provides information on the dynamics of the whole system
and Eqs.~(\ref{peak}) and (\ref{gr}) form the background for such
tunneling spectroscopy of quantum systems. Note, however, that the
subject of study, the spin-spin correlation function, is affected by
tunneling measurements (see the way the bare spin-spin correlation
function, Eq.~(\ref{bare}), was modified in the case of a single-spin
system, Eq.~(\ref{gamma})). We have shown how this modification may be
accounted for in the case of steady-state measurements of a single-spin
system. In a similar way it may be accounted for in more complicated
systems, and corrections to obtain the bare correlation function of the
system studied can, in principle, be made.

We note that we presented here a theoretical description of the
tunneling spectroscopy of a single localized spin-1/2 system. However,
such a description is also valid for any two-level system (quantum
dot), and Eq.~(\ref{transl}) describes the  mapping between the
two-level system and the spin-1/2 system.

In conclusion, we described the characteristics of the tunneling
current via an isolated single spin 1/2 (two-level system) in the
steady state, i.e., in the long time limit after switching on the
voltage or the tunneling matrix elements. We found optimal conditions
when the tunneling current carries maximum information about dynamic
quantum fluctuations of the localized spin. We showed what type of
information about the isolated spin may be extracted from measurements
of the tunneling current in the steady state.

	    We express our thanks to D.J. Scalapino, M.N. Kiselev, D.
Averin, A. Andreev, A. Shnirman, I. Martin, D. Mozyrsky, B. Spivak,
A.M. Tsvelik, and A.V. Balatsky for useful discussions. We acknowledge
A. Shnirman and Yu. Makhlin for sharing their results on the spin-spin
correlation function, Eq.~(\ref{tr1}), prior to publication. One of us
(M.H.) is grateful for the hospitality of the Los Alamos National
Laboratory. This work has been supported by the U.S. DOE.

\end{document}